\documentclass[11pt]{article}
\usepackage[utf8]{inputenc}
\pdfoutput=1
\usepackage{amssymb,amsmath,mathrsfs,enumerate}
\usepackage{graphicx,rotate,multicol}
\usepackage{float}
\usepackage{tocloft}
\usepackage{subcaption}
\usepackage[margin=10pt,labelfont=bf]{caption}
\usepackage{cite}
\usepackage{soul}
\usepackage[colorlinks=true,
			linkcolor=blue,
			urlcolor=blue,
			citecolor=teal]{hyperref}
\usepackage{multirow}
\usepackage{placeins}
\usepackage[normalem]{ulem}
\makeatletter
\let\Hy@linktoc\Hy@linktoc@page
\makeatother
\usepackage{color}
\definecolor{ourcolor}{rgb}{0.7, 0.25, 0.05}
\usepackage{tikz,braket}
\long\def\rpl#1!!#2!!{\textcolor{red}{#1} \textcolor{blue}{#2}}

\def \order(#1){{\mathcal O} \left(#1 \right)}

\evensidemargin=\oddsidemargin
\allowdisplaybreaks

\title{\color{black}{\bf High-redshift supermassive black hole population from core-collapse in self-interacting dark matter halos}}

\author {\bf Sambo Sarkar\footnote{sambos@iiserbpr.ac.in} \,and\, Ujjal Kumar Dey\footnote{ujjal@iiserbpr.ac.in}
	\\[10pt]
	\small\em Department of Physical Sciences, Indian Institute of Science Education and Research Berhampur,\\
	\small\em Ganjam, Odisha, 760003, India}
\date{}
\begin{document} 
	\maketitle
	
\abstract{Self-interactions between dark matter (DM) particles facilitate the inter-particle redistribution of energy within the central region of DM halos. Recent studies of dark matter spikes around massive black holes, and the diversity in rotation curves of dark matter rich low-mass galaxies motivate the exploration of  self-interaction cross-section $\mathcal{O}(10)\,\rm cm^2/gm$. At such high scattering rates, DM can lead to the formation of supermassive black hole seeds, through the gravothermal collapse of halo cores, under certain cosmological conditions. Quasars near cosmic dawn are powerful probes for examining the formation scenario of high redshift supermassive black holes, and their connection to structure formation. In this work we point out the favorable initial cosmological conditions that are likely to provide the black hole seeds resulting in supermassive black holes, mediated by core-collapse in self-interacting dark matter halos. Employing a semi-analytic prescription of core formation, and realistic mass accretion and halo merger histories, we compare the existing observations of high-redshift supermassive black holes with those predicted for our core-collapse framework. We find the median values of velocity-dependent self-interacting dark matter parameter space $\sigma/m_{\chi}\sim 56\, \rm cm^2/gm$, and $\omega=101\,\rm km/s$, assuming an Eddington accretion rate of unity. Sub-leading values are also presented for sub and super-Eddington accretion rates. We also report the derived self-interacting dark matter model parameters to account for the observed binned supermassive black hole mass functions at high-redshifts.}

\newpage

\hrule \hrule
\tableofcontents
\vskip 10pt
\hrule \hrule

\section{Introduction}

The existence of supermassive black holes (SMBHs) with masses $\sim 10^{10}\,M_{\odot}$ within the first billion years of the cosmic history pose a challenge to our existing understanding of cosmological structure formation \cite{Inayoshi:2019fun}. Since the discovery of quasars at redshifts $z>6$, by the Sloan Digital Sky Survey (SDSS) \cite{SDSS:2001emm,SDSS:2003iyw,Fan:2005es}, deeper optical and near-infrared surveys such as the Canada-France High-Redshift Quasar Survey (CFHQS) \cite{2010AJ....139..906W}, UKIDSS \cite{Lawrence:2006de}, Pan-STARRS1 \cite{Banados:2016mwo}, the VISTA Kilo-degree Infrared Galaxy Survey (VIKING) \cite{Venemans:2013npa}, the Dark Energy Survey (DES) \cite{DES:2017yqk}, Subaru's SHELLQs survey \cite{Matsuoka:2017frx,2019ApJ...880...77O,2021ApJ...908..235I}, and recent spectroscopic
compilations of XQR-30 \cite{DOdorico:2023sxe} have substantially increased the census of high luminosity quasars at high redshift $z >6$, with predicted black hole (BH) masses in the range, $M_{\rm BH}\sim(10^{8}-10^{10})\,M_\odot$. Quasars discovered beyond $z\sim7$ like ULAS J1120+0641 \cite{2011Natur.474..616M}, J1342+0928 \cite{Banados:2017unc}, and J0313-1806 \cite{2021ApJ...907L...1W}, demonstrate the existence of $\sim 10^{9}\,M_{\odot}$ BHs at an age of $\sim 700\,\mathrm{Myr}$. This rapidly growing sample of high-redshift quasars tend to severely constrain the BH formation scenarios. SMBHs observed at $z \geq 8$ must have formed and grown in less than 600 Myr, aided by an early BH seed, and sustained Eddington accretion \cite{Chen:2025vga}. However, realistic astrophysical environments with radiative feedback, finite gas supply, galaxy mergers make such sustained growth difficult \cite{2010A&ARv..18..279V,2014Sci...345.1330A,Inayoshi:2019fun}. The observational landscape has witnessed a dramatic increase, following the launch of James Webb Space Telescope (JWST). Deep spectroscopic surveys such as CEERS \cite{2025ApJ...983L...4F}, JADES \cite{2026ApJS..283....6E}, EIGER and FRESCO \cite{Matthee:2023utn}, RUBIES \cite{2025A&A...697A.189D}, PRIMER \cite{2025MNRAS.539.2685L}, and UNCOVER \cite{2024ApJ...974...92B} have also entered a previously inaccessible population of low-luminosity active galactic nuclei (AGN), over the redshift range $4\le z\le 9$. The discovery of Little Red Dots (LRDs) have revealed a population of compact red sources that may represent rapidly growing BHs embedded inside early-stage galaxies \cite{Matthee:2023utn,2023ApJ...954L...4K}. Collectively, these observations have substantially extended the census of accreting SMBHs toward the epoch of reionization, providing increasingly large samples for studying early BH growth, and its connections to structure formation \cite{2023ARA&A..61..373F}.\\

While primordial black holes (PBH) can form right after the Big Bang \cite{Carr:2009jm}, a variety of proposed SMBH formation scenarios require sufficient time. SMBH seeds may originate from the remnants of metal-free Population III stars (Pop III), which can produce light seeds with masses of order $(10^{1}-10^{3})\,M_\odot$ \cite{Madau:2001sc,Pelupessy:2007mt,Costa:2023xsz}. More massive seeds may arise from the collapse of supermassive stars formed through the rapid accumulation of pristine gas \cite{1993MNRAS.263..168H,Begelman:2006db,Shlosman:2015wma,Latif:2016qau}, or from runaway stellar collisions in dense nuclear star clusters \cite{AtakanGurkan:2003hm}. Pristine gas cloud collapse without fragmentation could yield direct-collapse black hole (DCBH) seeds with mass $M_{\rm seed} \sim (10^4 - 10^5) M_{\odot}$ \cite{Habouzit:2016nyf,2018MNRAS.476.3523L}. SMBH seeds may also be produced through
dark-matter capture in Pop III stars, which can induce their premature collapse into BHs
\cite{Ellis:2021ztw,Bhattacharya:2025dgx}. PBH formed in the early Universe provide an additional, non-stellar seed formation mechanism \cite{1974MNRAS.168..399C}. Given these SMBH seed formation channels, there remains a reasonable absence of consensus on the universally accepted explanation, for the observed mass range and population density of SMBHs, across the large redshift range of observations currently accessible. An alternative possibility for SMBH seed formation is through the gravothermal core collapse of self-interacting DM (SIDM) halos, that can produces a central BH at high redshifts \cite{Pollack:2014rja,Choquette:2018lvq}. Such a mechanism is particularly interesting because the SMBH seed formation timescales and mass are determined by the DM halo properties, its mass accretion history (MAH) and the DM self-interaction strength.\\

The standard cold dark matter (CDM) paradigm \cite{1984Natur.311..517B,Salucci:2018hqu}, together with the standard model (SM) of particle physics has been overall successful in explaining the large-scale structures of the universe \cite{DelPopolo:2002sz,Lisanti:2016jxe,Pace:2019vrs}. However, several long-standing discrepancies due to the mismatch between observations and $\Lambda$CDM predictions exist on the galactic scales. These include the discrepancy in observed galactic cores and cusps \cite{BoylanKolchin:2003sf,Baushev:2016pep}, the diversity in galaxy rotation curves\cite{Oman:2015xda,Kamada:2016euw}, and the too-big-to-fail issue \cite{BoylanKolchin:2011de,2014MNRAS.444..222G}. They have motivated extensions of CDM\footnote{These discrepancies may be artifacts of less sophisticated $N$-body simulations or less accurate inclusion of baryonic effects \cite{Vogl:2024ack,Martin-Alvarez:2022tfw}.} with velocity-dependent DM self-scattering \cite{Tulin:2017ara,Banerjee:2019bjp}. DM self-scattering aid in the transport of heat through the inner region of the halo. During the initial gravothermal evolution, heat flows inward from the periphery towards the core, and produces an approximately isothermal, lower-density core. After maximum core expansion, the heat flow reverses and the core starts to lose energy, thereby contracting. The core then heats up owing to its negative heat capacity, eventually entering the gravothermal collapse phase. \cite{Spergel:1999mh,Balberg:2002ue,Jun:2011jun,Tran:2025riw}. The endpoint of such catastrophic evolution may result in the formation of SMBH seeds. Since the characteristic seed masses produced through this mechanism are considerably larger than those expected from stellar remnants \cite{Arguelles:2023hab}, SIDM-induced core-collapse provides an attractive solution to the observed high-redshift SMBHs \cite{Balberg:2001qg,Jiang:2025jtr,Shen:2025evo}. Velocity-dependent DM self-interactions have been motivated to address diverse rotation curves in dwarf and low surface brightness galaxies (LSB) \cite{Zeng:2024xty,Roberts:2024uyw}, while simultaneously satisfying constraints from strong lensed galaxy clusters \cite{Sagunski:2020spe,Eckert:2022qia}. In addition, velocity-dependent DM self-interactions may address the formation of DM spikes around massive central BHs \cite{Alonso-Alvarez:2024gdz}. These studies however predict $\sigma/m\sim \mathcal{O}(10) \,\rm cm^2/gm$. At such high cross-sections, there arises the possibility for halo cores to develop gravothermal instability, and collapse to form BHs. Velocity-dependent DM self-interactions are therefore argued to produce sufficiently massive BH seeds \cite{Roberts:2024wup}. With core collapse in SIDM halos motivating the formation of SMBH seeds, the rapidly expanding observational census calls for a statistical comparison between observations and theoretical predictions, which require realistic descriptions of halo assembly histories, merger statistics and Eddington accretion rates. In this work, we model high-redshift SMBHs based on core collapse of velocity-dependent SIDM, and compare the observed SMBH population over the redshift interval $4\leq z\leq 11$. We extensively employ publicly available codes \texttt{SatGen} \cite{Jiang:2020rdj,Green:2021vkf} and \texttt{hmf} \cite{Murray:2013qza,Murray:2020dcd}, to generate the merger and accretion history of each host halo over its cosmological timescales. Additionally, we use the semi-analytical description of SIDM core formation discussed in \cite{Kaplinghat:2015aga}. We then report constraints on the DM halo properties such as mass $M_{\rm halo}$, concentration $c_{200}$, and the velocity-dependent SIDM parameters $(\sigma_0\,,w)$ within corresponding credible intervals, which provide an acceptable goodness-of-fit to the observed SMBH mass-redshift distribution, and binned mass density. We consider a broad range of host-halo mass $(10^7-10^{12})\, M_{\odot}$, and recent compilations of spectroscopic and broad-line AGN and quasars obtained from both pre-JWST and recent JWST surveys.\\

The paper is organized as follows. In section \ref{sec:seedvdsidm}, we discuss the numerical rationale of SMBH seed formation through velocity-dependent DM self-interactions. We then describe our workflow, which include the method for estimating the SIDM core, the initial cosmological setup essential to our core-collapse scenario, the BH merger dynamics, and the accretion mechanism of SMBH seeds in section \ref{sec:analysis}. Finally we present the results of our analysis in section \ref{sec:results}, by comparing the predictions of our SMBH mass and mass functions from the SIDM mediated core-collapse, with their observational counterparts. We then conclude and summarize our findings in section \ref{sec:conc}. 
\section{Black hole seeds from velocity-dependent self-interactions}
\label{sec:seedvdsidm}

High values of the specific self-interaction cross-section $\sigma/m_{\chi}$, can induce gravothermal instability within the central regions of SIDM halos \cite{Palubski:2024ibb}. An increase in the particle density at the core plunges the halo to a phase of core instability and subsequent collapse at $z_{\rm coll}$. These collapsed cores then become seeds for high-redshift SMBHs. The presence of baryons has been argued to speed-up core-collapse, with relatively weaker SIDM cross-sections \cite{Feng:2020kxv}. High-redshift observations have revealed more than 200 quasars, powered by accreting SMBHs \cite{2017ApJ...849...91M}. Masses of such BHs can reach $\sim 10^{10}M_{\odot}$ by a redshifts of $6.3$ \cite{Wu:2022njo}. Discoveries of such early SMBHs have reignited the debate on the formation and growth of BHs so large and so fast. The problem is further exacerbated by the fact that high-velocity recoils due to BH mergers in low-mass halos can induce velocity kicks up to $(500-1000)\,\rm km/s$, which can significantly reduce the efficiency of BH growth, by expelling them from the shallow potential wells of low-mass halos \cite{Volonteri:2006ma}. Some studies have tried to alleviate this problem by allowing BH to undergo short episodes of super-Eddington accretion \cite{Haemmerle:2020iqg}. Alternatively, SMBH seeds of masses $(10^4-10^6) M_{\odot}$ could also form via merging of compact stellar remnants in young galaxies with dense gaseous environment \cite{Boco:2020hhf}. This diversity motivates to look for possibilities beyond the existing framework. We therefore study the prospects of addressing SMBH seed formation using velocity-dependent DM self-interactions.\\

We consider the scenario in which self-interactions between fermionic DM can be mediated by scalars or gauge bosons \cite{Loeb:2010gj,Vogel:2012vol}. In the weakly-coupled perturbative limit or the Born regime, the differential cross-section, assuming Rutherford-type scattering is given by \cite{Ibe:2009mk,Robertson:2016qef,Yang:2022hkm},
\begin{equation}
\frac{\rm d\sigma}{\rm d\, cos \,\theta}=\frac{\sigma_0 \omega^4}{2[\omega^2+v^2\,\rm sin^2(\theta/2)]^2},
\label{eq:sigmadiff}
\end{equation}
where $\sigma_0$ is the specific DM self-interaction cross-section in the low-velocity limit, $\omega$ represents a characteristic velocity above which cross-section sharply declines, and $v$ is the relative velocity between the DM particles, $\theta$ is the scattering angle. For our purpose we use an effective cross-section $\sigma_{\rm eff}$, obtained by integrating $\rm d\sigma/ d\, cos \,\theta$ over the DM velocities and scattering angles \cite{Yang:2022hkm,Yang:2023jwn}. DM in halos are considered to follow a Maxwell-Boltzmann distribution \cite{Binney:1987bin}. However, statistical and empirical distributions have also been considered in the literature \cite{Maity:2020wic,Bose:2022ola}. Recent studies have shown that velocity-dependent DM self-interactions drive the halos through a sequence of core formation, followed by gravothermal collapse \cite{Nishikawa:2019lsc,Turner:2020vlf,Outmezguine:2022bhq}.
\section{Cosmology of SMBHs from core-collapse}
\label{sec:analysis}

Gravothermal core collapse in DM halos is sensitive not only to the velocity-dependent SIDM parameter space, but also to the initial cosmological conditions and MAH. We consider our initial cosmology from high-redshifts $z\sim 20$, consistent with the $\Lambda$CDM power spectra and halo mass functions \cite{Sheth:1999mn,Reed:2006rw,2013ApJ...770...57B}. Therefore, a discussion on the cosmological and astrophysical setup is now in order. In this section we briefly outline the steps followed during core formation, halo mass accretion, collapse of the SIDM halo cores into BHs, merger dynamics and Eddington growth of BHs.
\subsection{Core formation in SIDM halos}
\label{subsec:coreformation}

Self-interactions between DM induce an initial thermalization at the center of DM halos, where particles tend to follow an isothermal distribution \cite{Burger:2018sqp,Robertson:2020pxj}. The core size $r_1$, and core density $\rho(r_1)$, of this region is found to be correlated with the strength of DM self-interaction \cite{Elbert:2014bma,Kamada:2019wjo,Ray:2022ydr,Jiang:2022aqw,Ray:2025xrv}. As we study the formation of SMBH seeds from the collapse of halo cores, an a priory understanding and estimation of the core radius and mass enclosed within it becomes essential. We follow the prescription of \cite{Kaplinghat:2015aga}, which determines the core radius $r_1$ and the mass enclosed $M_{\rm core}$, by a semi-analytic formalism governed by
\begin{equation}
\rho(r_1)\, \frac{ \left\langle \sigma \,v_{\rm rel} \right\rangle}{m_{\chi}}\,t_{\rm age} \sim 1,
\end{equation}
where the symbols carry their usual meaning with $t_{\rm age}$ denoting the cosmological timescale of halo evolution. The average is taken over the product of velocity-dependent SIDM cross-section and the relative velocity between DM particles. The condition states that each DM particle inside $r_1$ scatters at least once within a time frame of $t_{\rm age}$. DM inside the core therefore follows an isothermal distribution \cite{Binney:1987bin,Kaplinghat:2015aga}, whereas outside $r_1$ DM particles are distributed according to the NFW profile \cite{Navarro:1995iw}. For large values of $\sigma/m_{\chi}$ relevant for core-collapse, the core acts as an optically dense medium of boundary radius $r_1$. Within $r_1$, DM gets scattered into the center, instead of the halo periphery. This propels the core into a phase of gravitational instability where the core density starts to grow, triggering a spontaneous collapse. The dynamics of core-collapse depends on the DM halo mass $M_{\rm halo}$, concentration $c_{200}=r_{200}/r_{\rm s}$, and also the epoch at which the progenitor halos start to virialize $z_{\rm vir}$. Here $r_{200}$ is the virial radius of a DM halo of mass $M_{\rm halo}$. The timescales required for the core to collapse is given by \cite{Balberg:2002ue,Pollack:2014rja,Essig:2018pzq,Jiang:2025jtr},
\begin{equation}
t_{\rm cc}\sim \frac{150}{C} \frac{1}{(\sqrt{4\pi G\rho_{\rm s}}) \,\rho_{\rm s} r_{\rm s}\sigma_{\rm eff} },
\label{eq:tcc}
\end{equation}
\begin{figure*}[t!]
	\begin{center}
		\subfloat{\includegraphics[scale=0.3]{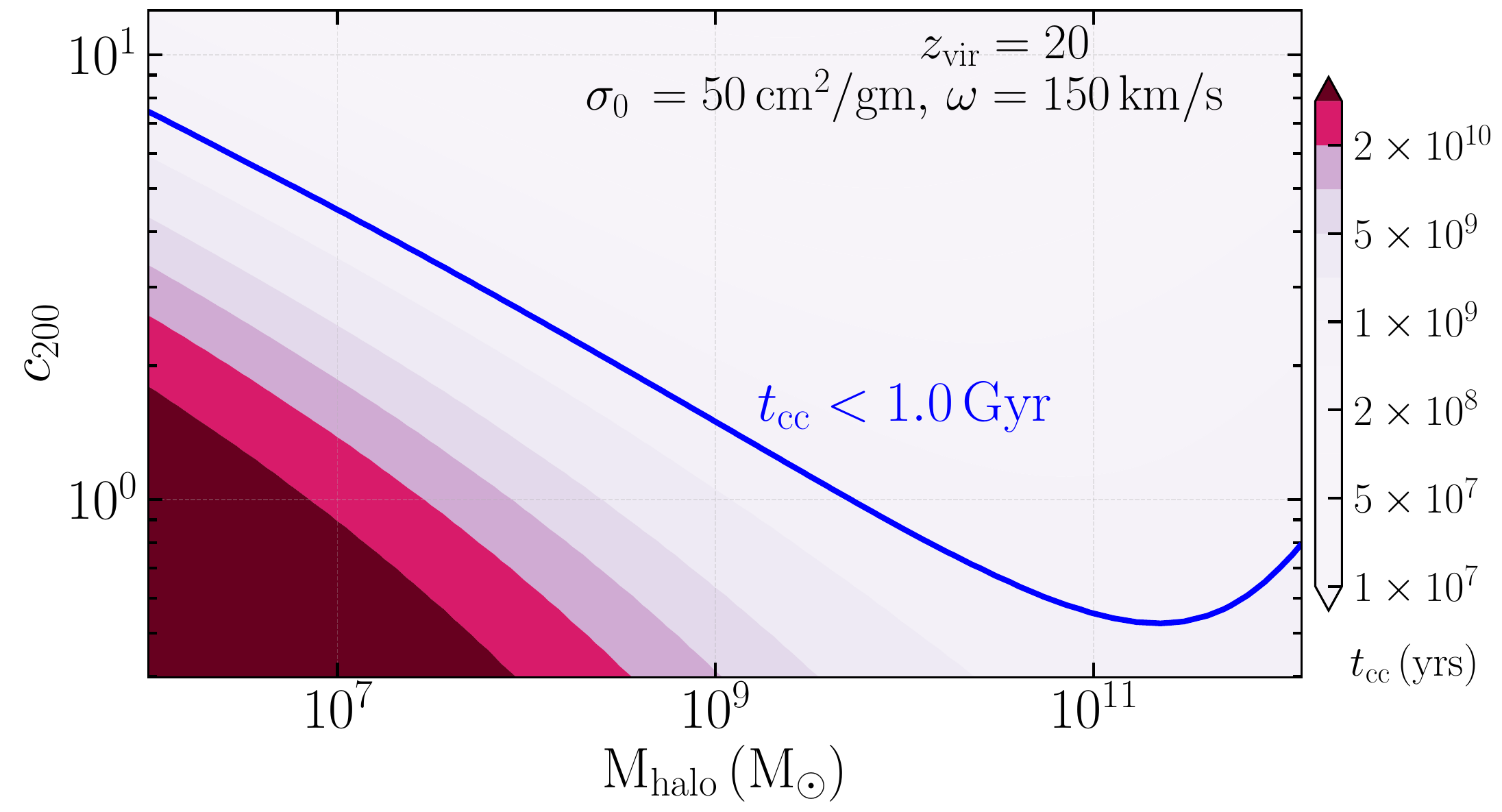}}
		\caption{Contour plot for DM halo mass vs. halo concentration parameterized over the core-collapse timescale $t_{\rm cc}$. The benchmark values of $\sigma_0/m=50\,\rm cm^2/gm$, and $\omega = 150\,\rm km/sec$, with a halo formation redshift $z_{\rm vir}$ of 20.}
		\label{fig:corecollapse}
	\end{center}
\end{figure*}
where $\rho_{\rm s}$ and $r_{\rm s}$ are the NFW model parameters, with $C=0.75$, calibrated with $N$-body simulations \cite{Jun:2011jun,Essig:2018pzq,Nishikawa:2019lsc}. The effect of velocity-dependent DM self-interactions enters the above equation through the effective scattering cross section $\sigma_{\rm eff}$, discussed in section \ref{sec:seedvdsidm}. Consequently, the earliest forming and most massive progenitors of each halo provide the most favorable environments for the growth of SMBH seeds. However, the halo parameters $(M_{\rm halo}\,,c_{200}$ and $z_{\rm vir}$) being highly correlated \cite{Ludlow:2013vxa,Diemer:2014gba}, an understanding of the progenitor evolution becomes necessary. There have been studies to understand this correlation using both $N$-body simulations \cite{Wang:2022spb}, and semi-analytic methods \cite{Correa:2015dva}. In order to see the correlation between these initial cosmological parameters, and their SIDM counterparts, in figure \ref{fig:corecollapse}, we plot the contours for core-collapse timescales over the halo mass and halo concentration for the benchmark choices of $z_{\rm vir}=20$, $\sigma_0/m=50\,\rm cm^2/gm$, and $\omega = 150\,\rm km/s$. We see that high concentration halos are more likely to collapse under the SIDM paradigm within timescales less than a Gyr. The formation of high-redshift SMBHs in this scenario becomes challenging as all progenitor halos, particularly that of low-mass halos, might not have the required concentration for reliably fast BH formation. In order to establish the correlation between the cosmological halo and SIDM parameters, we model the MAH of each host halo and the evolution of their cores in all the individual progenitor halos. This gives us the opportunity to examine their likelihood to form SMBH seeds.
\subsection{Mass accretion and halo mergers}
\label{subsec:inicosmo}

Halos at very high redshifts feature rampant merging and therefore are often not in strict equilibrium \cite{2011MNRAS.416..242D}. In order to accurately account for mergers and MAH, we use the publicly available code \texttt{hmf} to generate the host halo number densities at redshifts of interest. We extensively use \texttt{SatGen} to simulate the MAH of every DM halo and trace their progenitor branches back to $z_{\rm vir} \sim 20$. The semi-analytic modeling of \texttt{SatGen} is based on the Extended Press-Schechter formalism \cite{1974ApJ...187..425P,1993MNRAS.262..627L}, and the algorithms discussed in \cite{Jiang:2013kza,Benson:2016pht}. In \texttt{SatGen}, we set the minimum resolution of the progenitor halo at $\sim (10^4-10^6)\,M_{\odot}$.
\begin{figure*}[t!]
	\begin{center}
		\subfloat{\includegraphics[scale=0.36]{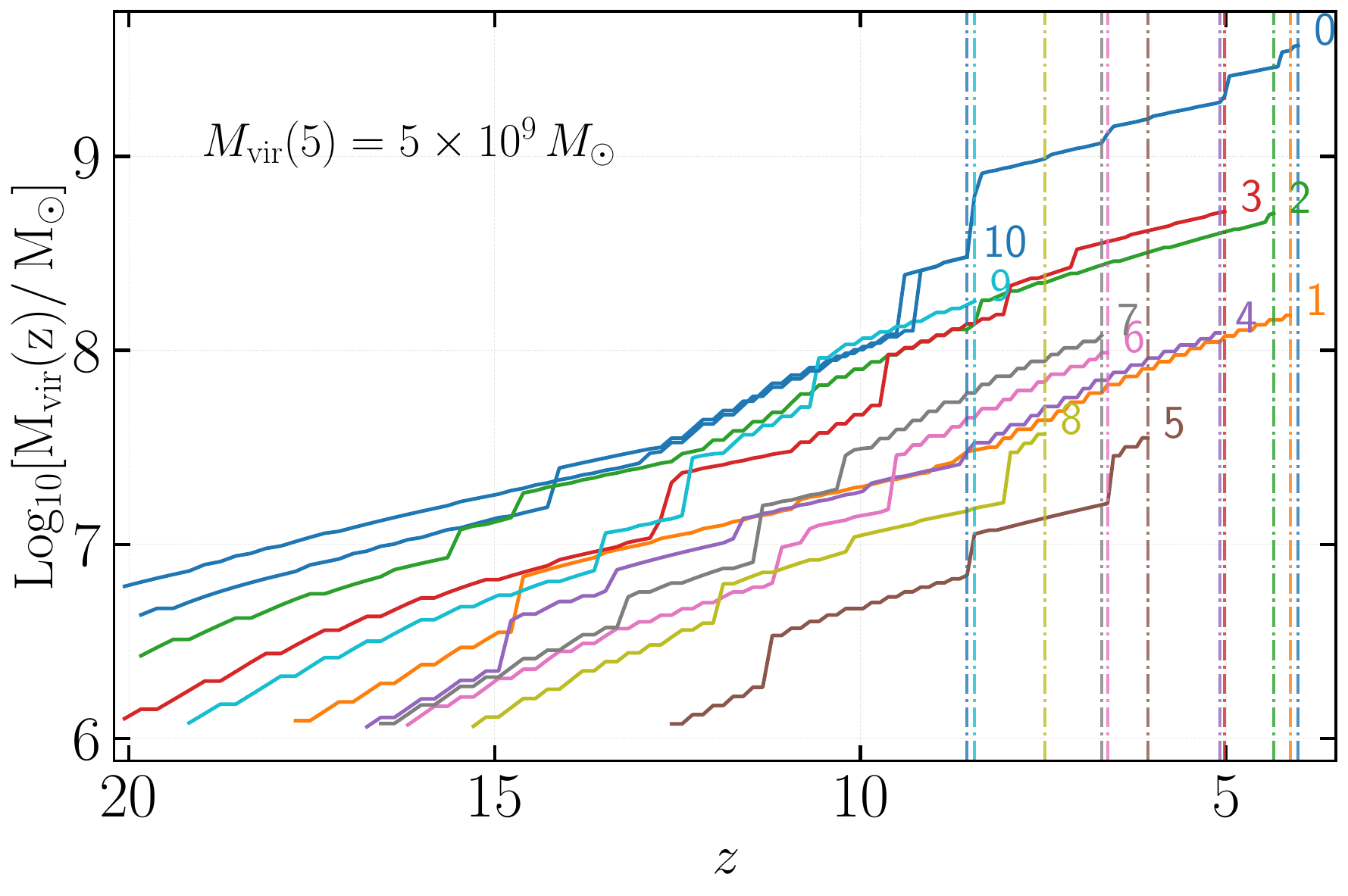}}
		\caption{Mass accretion and merger history of a $5\times 10^9\,M_{\odot}$ DM halo generated using \texttt{SatGen}. The first 11 of the most massive progenitors, extending to $z_{\rm vir}\sim20$, with the primary branch identified with the $0^{th}$ index are shown. The vertical dashed-dotted lines indicate the redshift at which the progenitors merge with their massive counterparts, generating larger halos in succession.}
		\label{fig:MAH}
	\end{center}
\end{figure*}
In figure \ref{fig:MAH} we show the MAH for a halo of mass $\sim 5 \times 10^{9}\,M_{\odot}$, generated using \texttt{SatGen}. The primary progenitor halo (identified by the index 0) extends to $z_{\rm vir}\sim 20$, along with few of the successive massive progenitor halos, showing their MAH and mergers with redshift.\\

Each progenitor-branch data file generated using \texttt{SatGen} contains the halo mass and concentration as functions of redshift. We then model the progenitor halos at each redshift using the semi-analytic description discussed in section \ref{subsec:coreformation}, and trace the evolution of progenitor halo parameters ($r_1,\,M_{\rm core},\,r_{\rm s}$, and $\rho_{\rm s}$) with redshift. This allows us to gauge the redshifts at which the core maximally expands, $z_{\rm core}$, and then collapses to form a BH at  $z_{\rm coll}$. We determine the mass and extent of the core at $z_{\rm coll}$, which provides us with a realistic estimate of the initial BH seed mass $M_{\rm seed}$ and its formation redshift $z_{\rm seed}$, from the simulations itself. Here we consider $z_{\rm coll} \sim z_{\rm seed}$. We find that more than one progenitor halo can host early forming BHs, depending on the SIDM and progenitor halo parameters. These BHs hence formed merge along with their progenitor halos, to generate more massive BHs. The merger of BHs inside their host halos are delayed due to the dynamical friction of the intervening DM medium given by \cite{Boylan-Kolchin:2007bvo},
\begin{equation}
t_{\rm DF}= 0.216 \frac{(m_{\rm r})^{1.3}}
{\ln\left(1+m_{\rm r}\right)} e^{1.9\eta_{\rm orb}} x_{\rm circ} t_{\rm cross}.
\label{eq:merge}
\end{equation}
Here $m_{\rm r}$ is the ratio of virial masses for the two merging progenitor halos, $\eta_{\rm orb}$ is the orbital circularity, $x_{\rm circ}$ characterizing the orbital energy, and $t_{\rm cross}=\sqrt{3/(4\pi G\Delta\rho_{\rm crit})}$ is the virial crossing time. The values of which are adopted from \cite{Zentner:2005wh,Jiang:2025jtr}. Here $\Delta=200$, and $\rho_{\rm crit}$ is the critical density of the universe. For each merging pair, the virial masses of the host halos are evaluated from their respective MAHs at $z_{\rm merge}$. Here $z_{\rm merge}$ is the redshift at which the individual host-halos merge. Gravitational waves can be emitted as a result of such mergers at high redshifts, which can also be used to study the SIDM parameter space. We leave this for future explorations. As and when a progenitor halo is found to undergo core-collapse, satisfying the condition $t_{\rm cc} \leq t(z_{\rm merge})-t(z_{\rm vir})$, we consider them to host a BH seed. Subsequently, a BH thus formed is consider to merge with its massive counterparts if $t_{\rm DF} + t(z_{\rm merge}) \leq t(z_{\rm obs})$. These seeds then evolve in near-Eddington accretion rates $\eta \sim 1.0$, over a period of $t(z_{\rm obs})-t(z_{\rm coll})$ or  $t(z_{\rm merge})-t(z_{\rm coll})$, depending on the progenitor branch.
\subsection{Eddington accretion of BH seed}
\label{subsec:eddington}

After the SMBH seed is formed, it starts to accrete the surrounding gas at a rate decided by the outward radiation pressure from the BH ejecta, and the inward gravitational gas pull. The Eddington accretion rate, for the duration of $t_{\rm obs} - t_{\rm seed}$, is given by \cite{2013fgu..book.....L},
\begin{equation}
M_{\rm BH}^{\rm obs} = M_{\rm Seed} \times \exp\left[\frac{\eta(1-\epsilon_r)\left(t_{\rm obs}-t_{\rm seed}\right)} {t_{\rm Sal}} \right].
\label{eq:eddington}
\end{equation}
Here $t_{\rm obs}$, and $t_{\rm seed}$ are the cosmic times at which the SMBH is observed, and the SMBH seed is formed. As a conservative limit we consider that the BH can accrete the ambient gas till $t_{\rm obs}$. Here $\eta$ denotes the accretion efficiency which is taken to be 1.0 for Eddington accretion scenario, 0.6 for sub-Eddington and 1.4 for super-Eddington in this work. The corresponding Salpeter time $t_{\rm Sal} =45.1(\epsilon_r/0.1)$ Myr \cite{1964ApJ...140..796S}, where the radiative efficiency $\epsilon_r$ is taken to be 0.1, which depends on the BH angular momentum \cite{Shapiro:2004ud}. The Eddington accretion sets the phase for rapid exponential growth of the BH seed. The accretion onto the central SMBH seed continues until a dynamical equilibrium between the SMBH seed and the baryons in host-halo is reached.
\section{Comparing with quasar observations}
\label{sec:results}

After evolving the halos from a redshift of 20 to 4, we now have an assembly of SMBH mass with redshift, for $\sigma_0/m$ in the range of $(10-100)\,\rm cm^2/gm$, and $\omega$ in the range of $(50 - 300)\,\rm km/s$. In this section, we compare the results of our generated SMBH catalog, to the existing observational data set of SMBH mass and distribution functions. We present our results on the SIDM model parameters, and the DM halo mass, specific to our SMBH formation mechanism.
\subsection{Predicted SMBH mass with redshift}
\label{subsec:results1}

\begin{figure*}[t!]
	\begin{center}
		\subfloat{\includegraphics[scale=0.405]{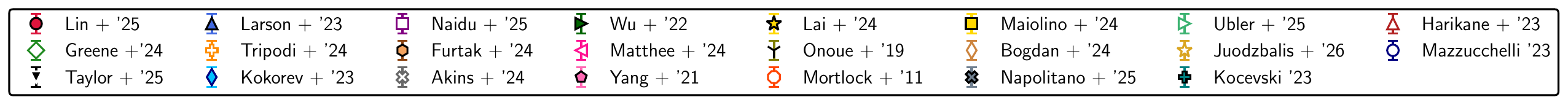}}\\
		\subfloat{\includegraphics[scale=0.19]{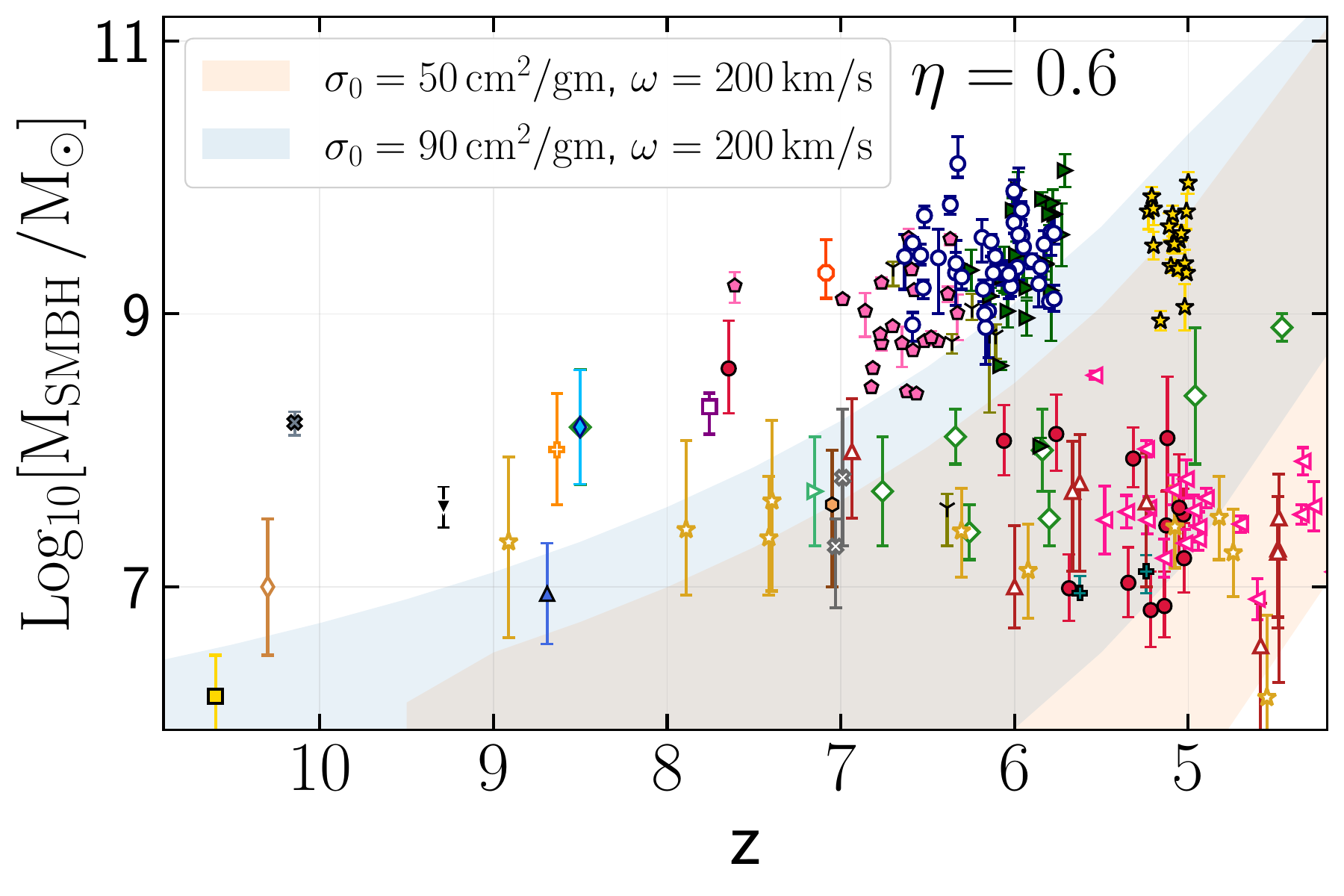}}
		\subfloat{\includegraphics[scale=0.19]{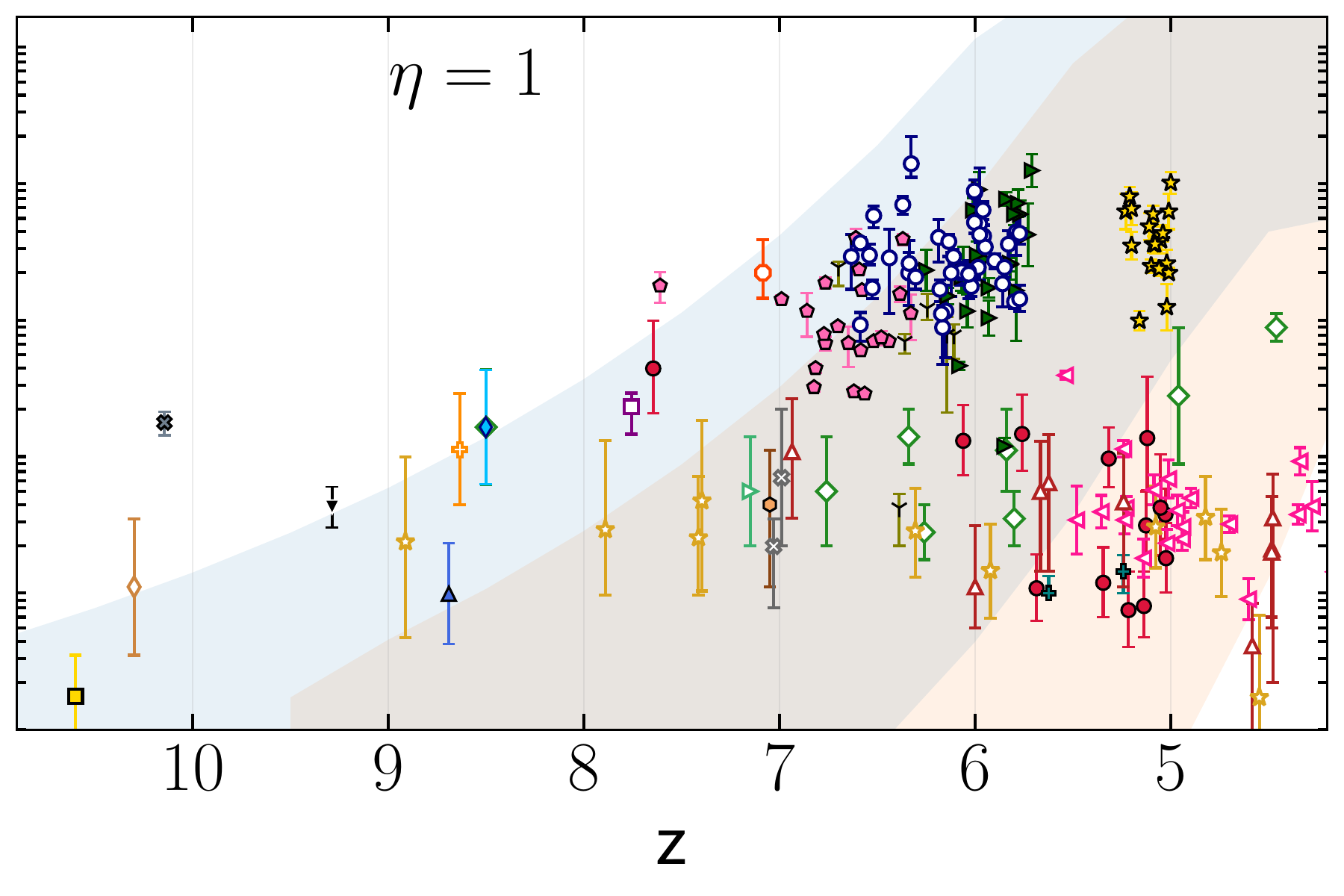}}
		\subfloat{\includegraphics[scale=0.19]{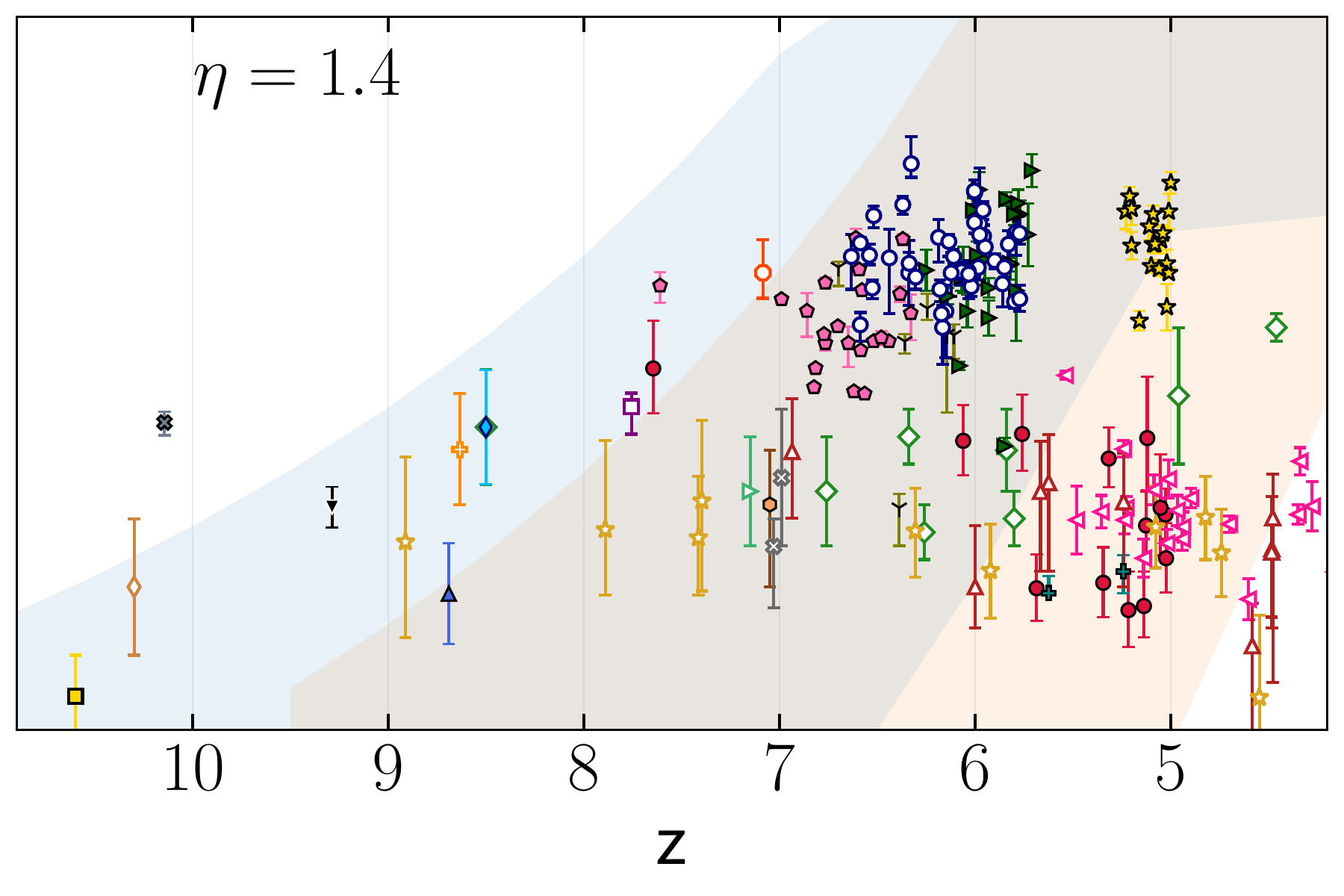}}
		\caption{Variation of observed and predicted SMBH mass with redshift. The yellow and blue regions denote the predicted SMBH mass band within $(16-84)^{\rm th}$ percentile, for the benchmark values of $\sigma_0$ = 50 and 90 $\rm cm^2/gm$, respectively, from core-collapse in SIDM halos, with $\omega=200\,\rm km/sec$. The data points along with the error bars denote observations of high redshift quasar ($4\leq z \leq 11$) measurements \cite{2011Natur.474..616M,2019ApJ...880...77O,Yang:2021imt,Wu:2022njo,2023A&A...676A..71M,Maiolino:2023zdu,Bogdan:2023ilu,2023ApJ...953L..29L,2023ApJ...957L...7K,Matthee:2023utn,2023ApJ...959...39H,2023ApJ...954L...4K,2024MNRAS.531.2245L,2024Natur.628...57F,2024ApJ...964...39G,2025NatCo..16.9830T,2025ApJ...989...75N,Naidu:2025rpo,2025arXiv250921575U,2025ApJ...989L...7T,2026ApJ...996...93L,2025ApJ...991...37A,2026MNRAS.546ag086J}.}
		\label{fig:CompMvsz}
	\end{center}
\end{figure*}
Observations have validated the presence of SMBHs within the first billion year of our cosmic history \cite{2023ARA&A..61..373F}. In the redshift range 5 to 7, most reliable SMBH mass estimates come from broad-line quasars which combine the width of a broad emission line with the continuum luminosity of the quasar \cite{Fan:2005es,2011Natur.474..616M,2015Natur.518..512W}. For such measurements the Mg{\,\sc ii} line is preferred, as estimates based on C{\,\sc iv} have larger systematic uncertainties \cite{2016MNRAS.461..647C}. Traditional quasar surveys preferentially detect the brightest and hence the most accreting BHs, thereby introducing strong selection effects in the data set. At a higher redshift of 7.54, quasar J1342$+$0928 is estimated to contain a SMBH of mass $\simeq8\times10^{8}\,M_\odot$ \cite{Banados:2017unc}, and J0313$-$1806 at $z=7.642$ is estimated to host a SMBH of mass $(1.6\pm0.4)\times10^{9}\,M_\odot$ \cite{2021ApJ...907L...1W}. These observations establish that SMBHs with mass $\geq10^{9}\,M_\odot$ existed at $t_{\rm age }\lesssim 0.7\,{\rm Gyr}$ after the Big Bang, and place strong constraints on the SMBH seed mass, seed formation redshifts, and the accretion rate. The XQR-30 programme obtained high signal-to-noise from the VLT-XSHOOTER spectra for a large sample of quasars at $z\gtrsim6$, and derived their BH masses and accretion rates \cite{DOdorico:2023sxe}. By studying $42$ luminous quasars they found SMBHs of mass $(0.8-12)\times10^{9}\,M_\odot$ \cite{2023A&A...676A..71M}. Six quasars within $6.1<z<6.7$, inferred from Mg{\,\sc ii}, based the SMBH masses in the range $(10^{7.6}-10^{9.3})\,M_\odot$ \cite{2019ApJ...880...77O}. The quasar SDSS J0100$+$2802 at $z=6.30$ hosts an exceptionally massive BH with mass $10^{10}\,M_\odot$ \cite{2015Natur.518..512W}. CEERS$\_1019$ at $z=8.679$ shows a broad H$\beta$ emission component that has been interpreted as AGN, the inferred BH mass corresponds to $9\times10^{6}\,M_\odot$ \cite{2023ApJ...953L..29L}. UHZ1, a gravitationally lensed galaxy at $z\sim 10.1$ is estimated to host a rapidly accreting BH with inferred mass $(10^{7}-10^{8})\,M_\odot$ by X-ray observations \cite{Bogdan:2023ilu}. At a higher redshift of 10.603, GN-z11 has been argued to host a SMBH of mass $\sim10^{6}\,M_\odot$ \cite{Maiolino:2023zdu}. The combination of luminous quasars from wide-field surveys, and faint broad-line AGN from JWST therefore provides a substantially broader sampling of the SMBH population.\\
\begin{figure*}[t!]
	\begin{center}
		\subfloat[\label{sf:chisq1}]{\includegraphics[scale=0.19]{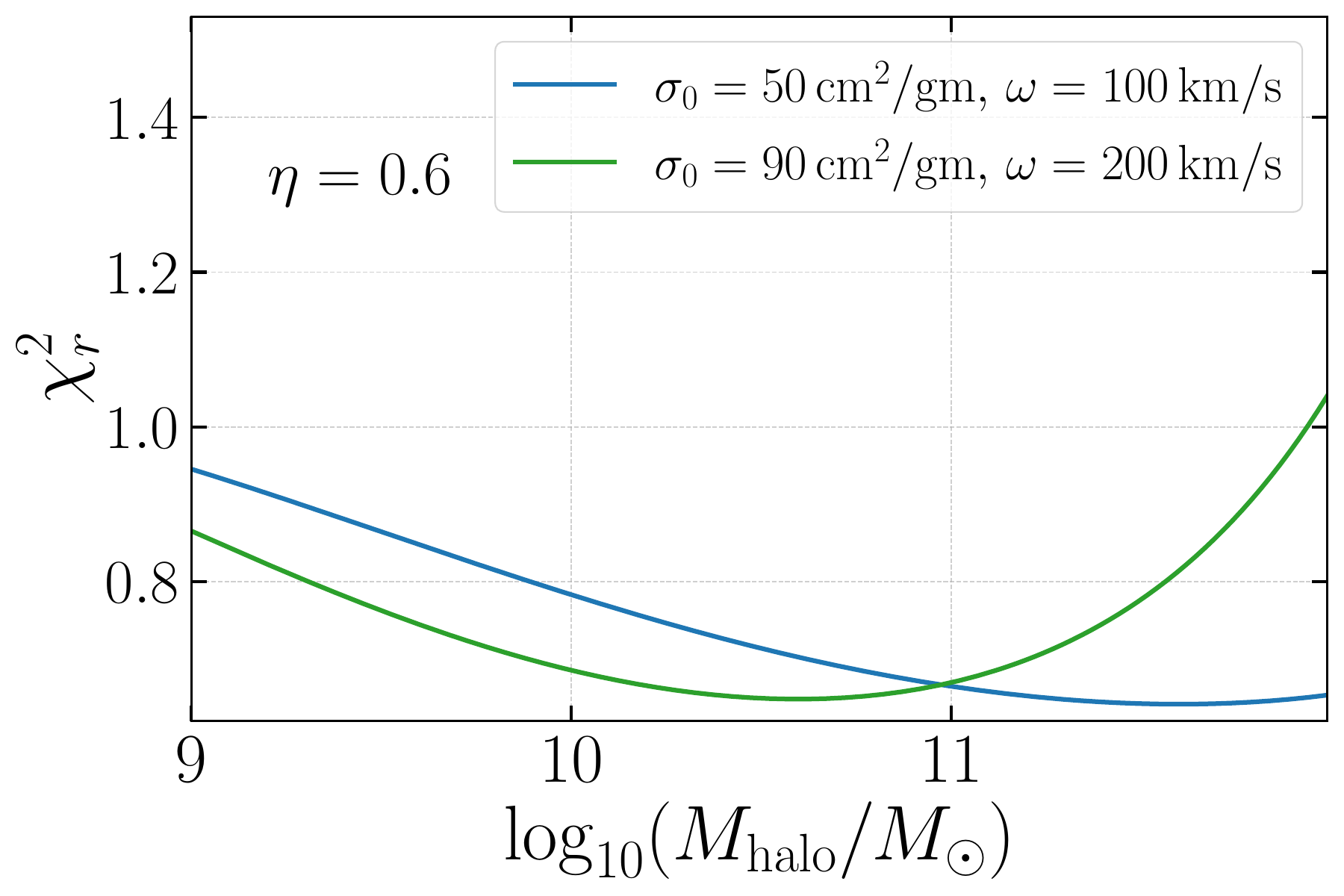}}
		\subfloat[\label{sf:chisq2}]{\includegraphics[scale=0.19]{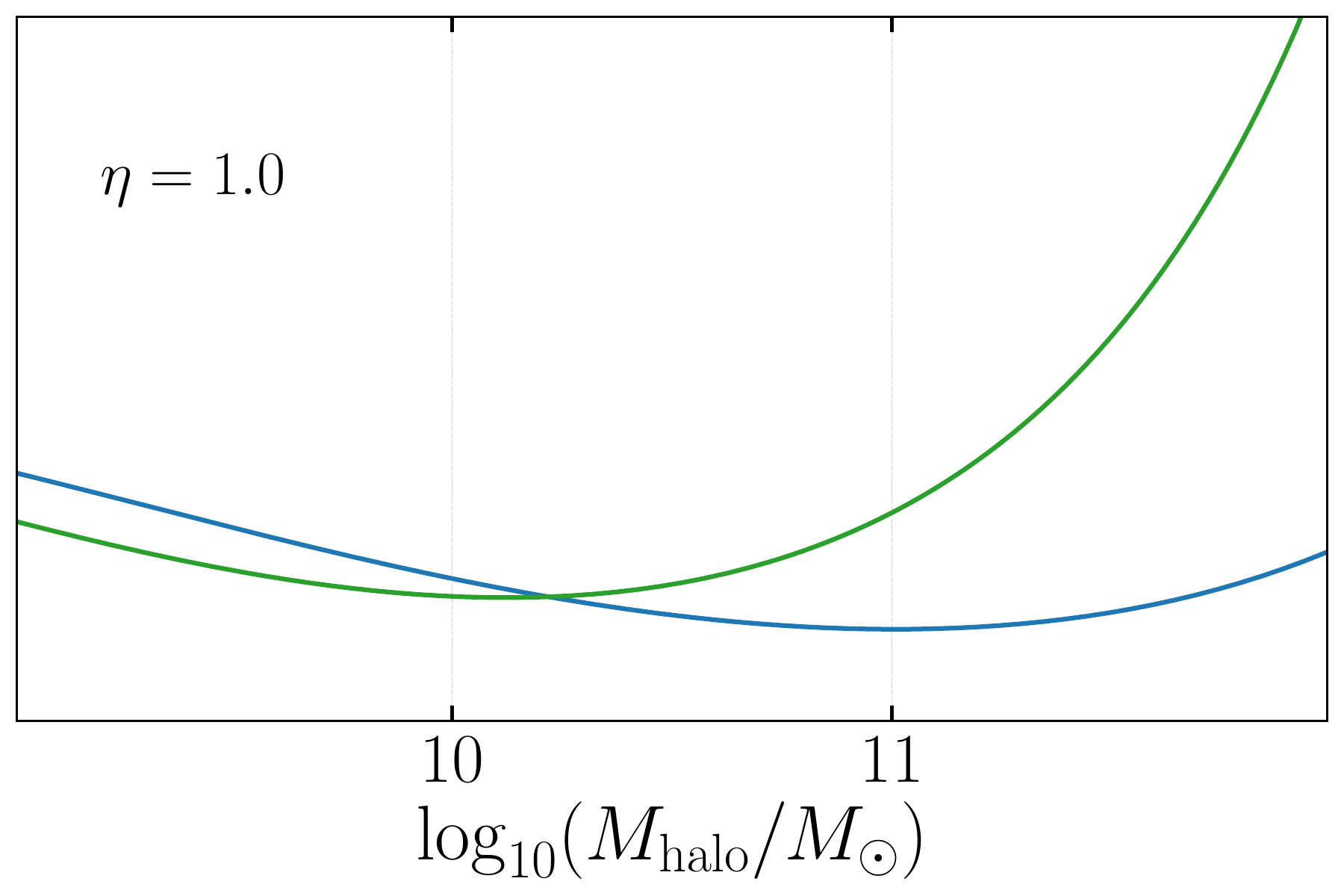}}
		\subfloat[\label{sf:chisq3}]{\includegraphics[scale=0.19]{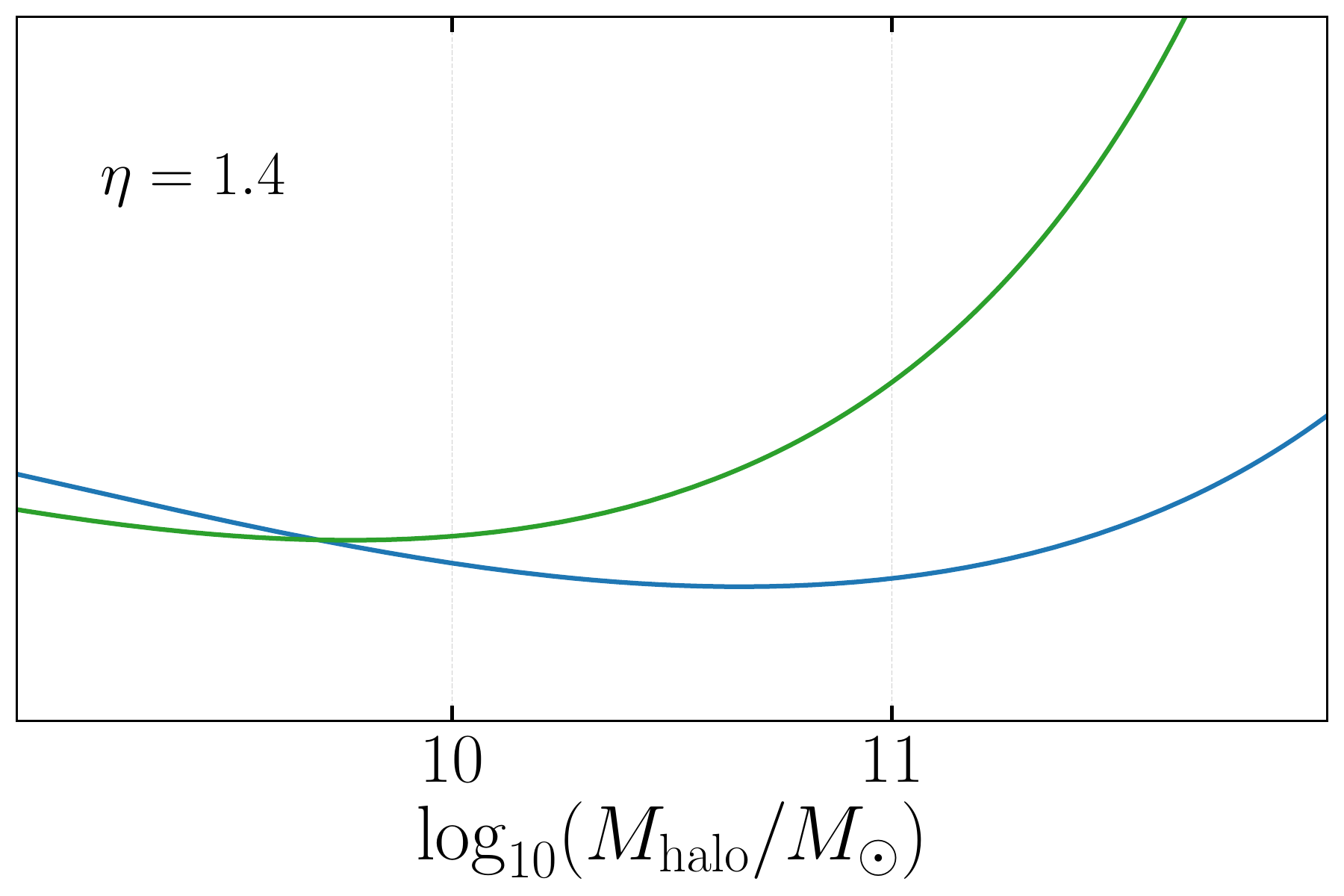}}\\
		\subfloat[\label{sf:chisq4}]{\includegraphics[scale=0.19]{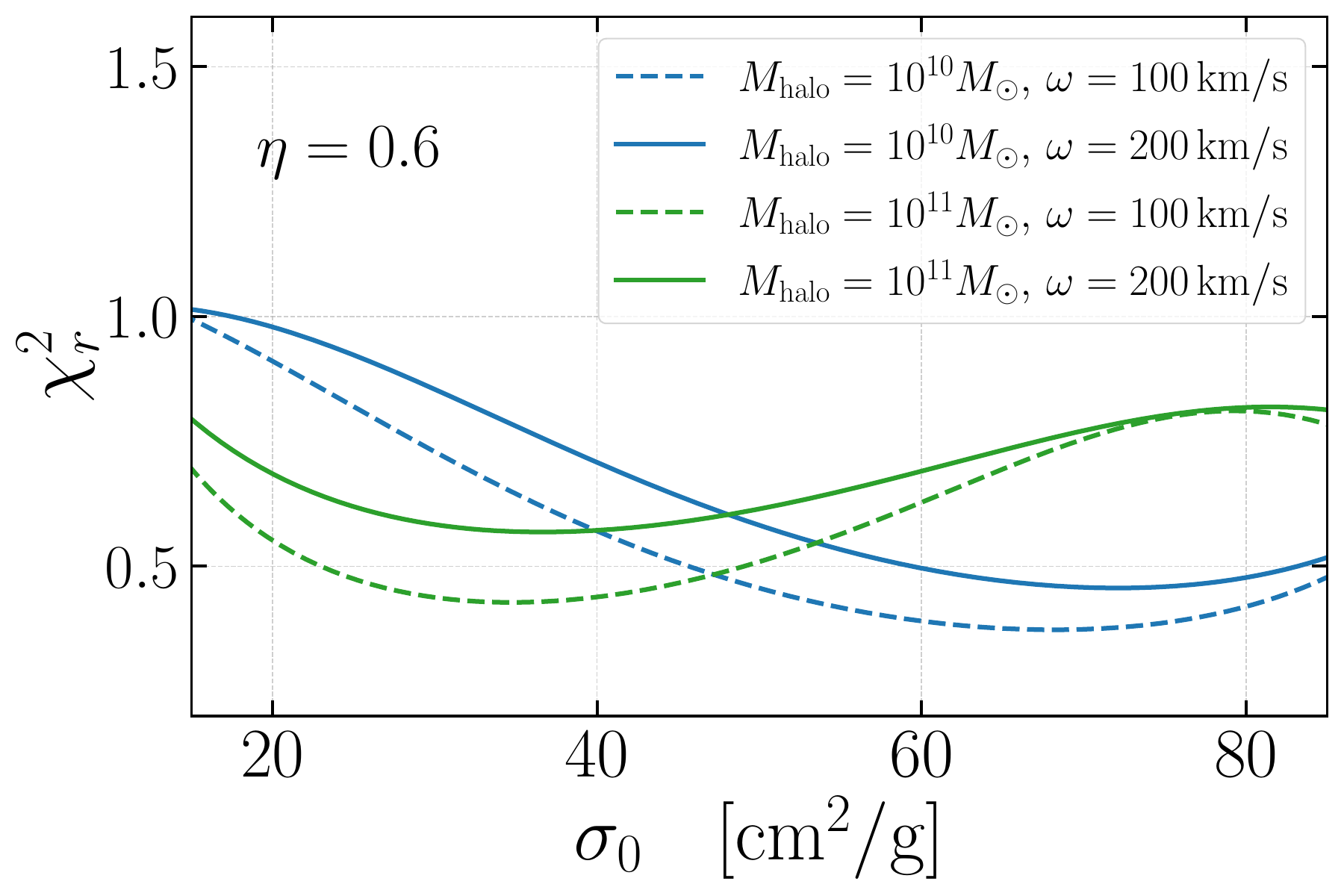}}
		\subfloat[\label{sf:chisq5}]{\includegraphics[scale=0.19]{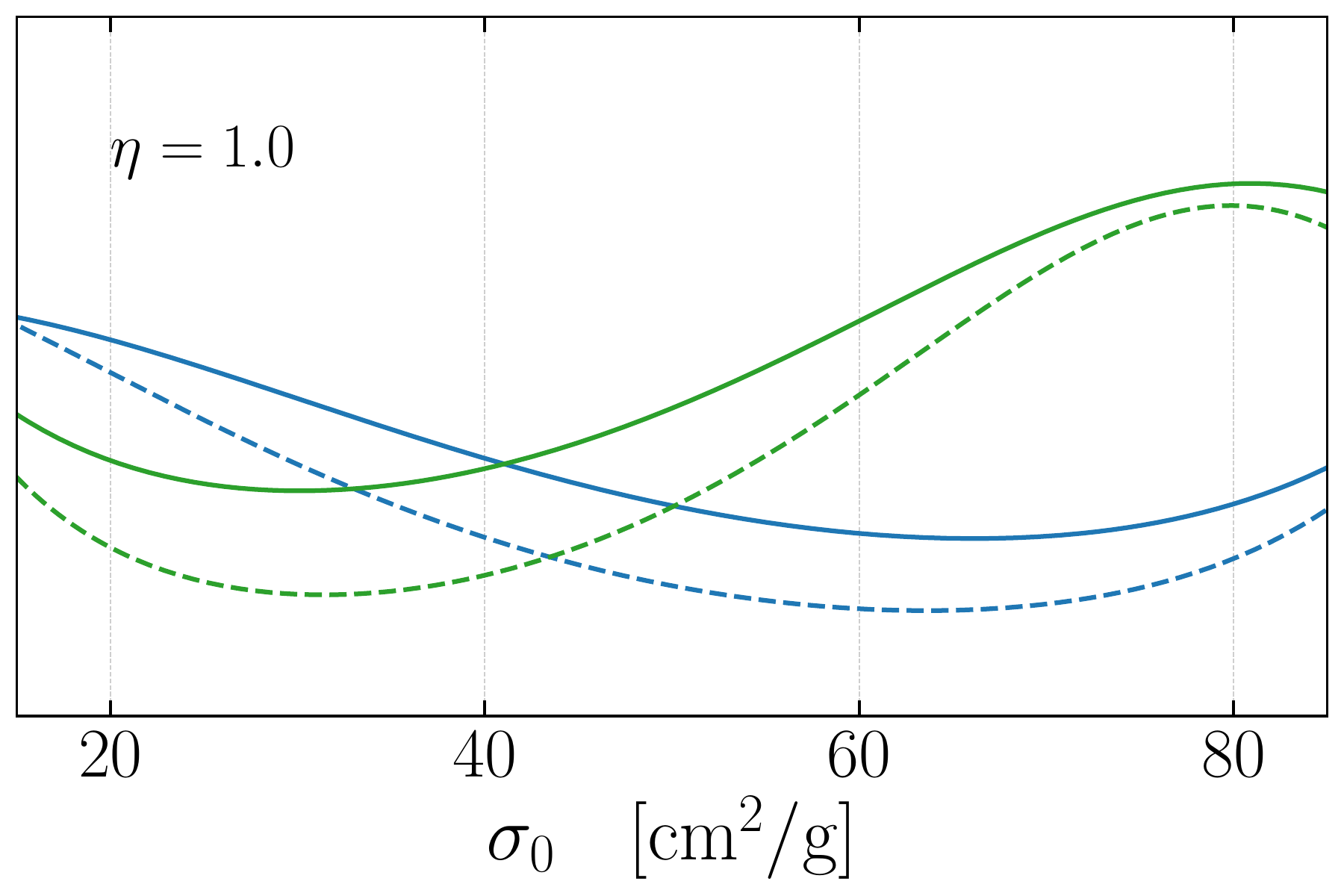}}
		\subfloat[\label{sf:chisq6}]{\includegraphics[scale=0.19]{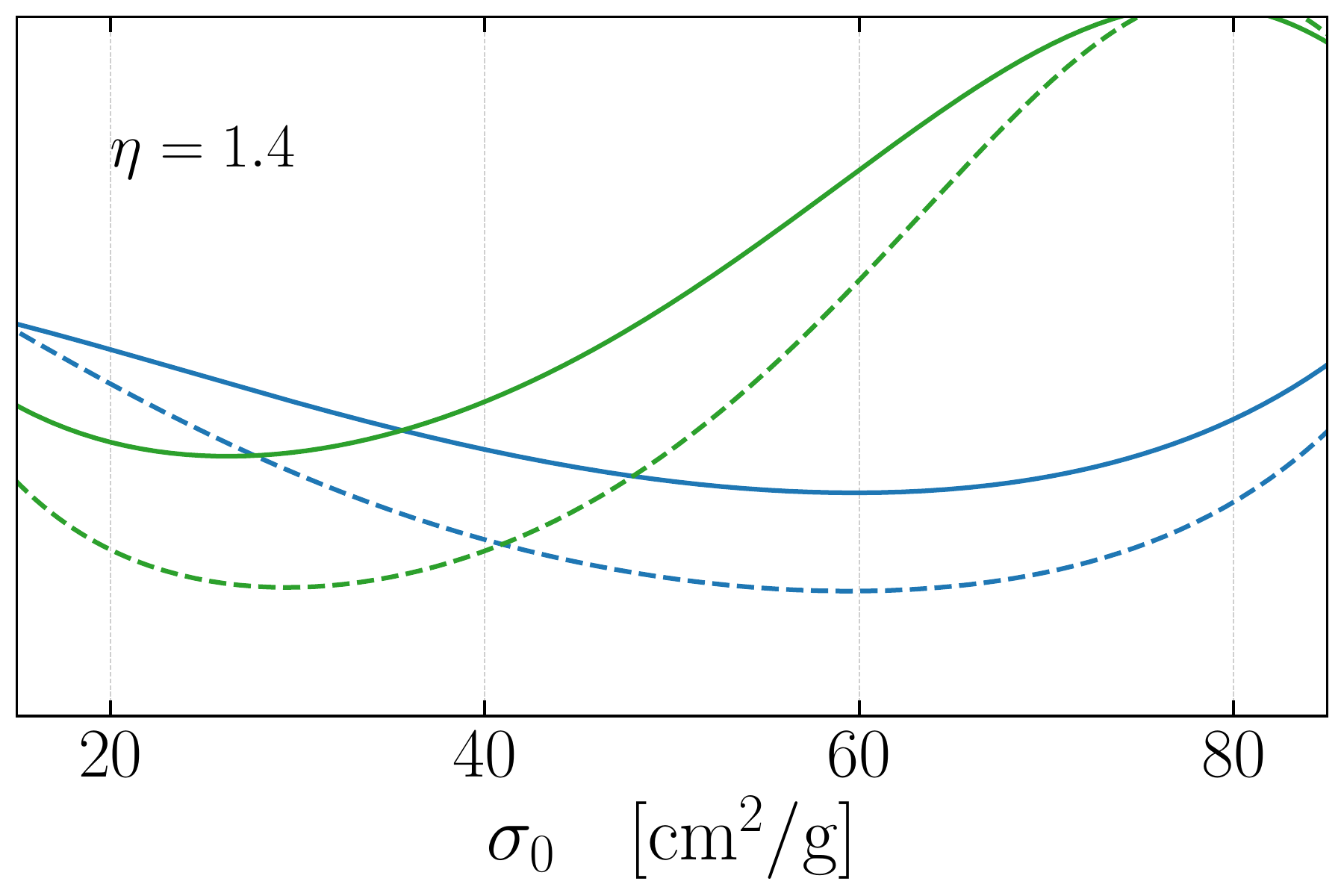}}
		\caption{\textit{top}: $\chi_r^2$ vs. DM halo mass, parameterized over two benchmark values of SIDM parameters $\sigma_0=50\,\rm cm^2/gm,\, \omega=100\, \rm km/s$ and $\sigma_0=90\,\rm cm^2/gm,\, \omega=200\, \rm km/s$. \textit{bottom}: $\chi_r^2$ vs. $\sigma_0$, parameterized over the DM halo mass. The left, middle, and right panels are for Eddington accretion rates of 0.6, 1.0, and 1.4 respectively.}
		\label{fig:Chisq}
	\end{center}
\end{figure*}

The observed redshifts are determined from spectroscopy, generally using the systemic or broad emission-line spectrum. The SMBH mass is inferred from the properties of H$\beta$ broad-line region (BLR). For distant quasars, the dominant technique is the single-epoch virial (SEV) method, which assumes that the BLR gas is gravitationally bound to the central BH. However, the SEV mass estimates also have systematic uncertainties of typically several tenths of a dex \cite{2012ApJ...753..125S,Chaves-Montero:2021bsz}. For $z>6$ quasar observations are most commonly estimated, based on the UV Mg{\,\sc ii} ($\lambda2798$) and C{\,\sc iv} ($\lambda1549$) lines, combined respectively with the continuum luminosity at approximately 3000 and 1450 \AA\ \cite{McLure:2002um,2012ApJ...746..169S}. With JWST-NIRSpec, the Balmer lines H$\beta$ and H$\alpha$ have become accessible and can be used in analogous virial estimators. The JWST samples use broad H$\alpha$ or H$\beta$ line widths, together with luminosity-based BLR radius relations to infer $M_{\rm BH}$ \cite{2023ApJ...959...39H,2026Natur.653.1017J}. At $z=8.50$,  UNCOVER-20466 has an inferred SMBH mass $\sim10^{8}\,M_\odot$ \cite{2023ApJ...957L...7K}. The strongly lensed source Abell 2744-QSO1 at $z=7.045$, identified a broad H$\beta$ emission, implying a BH of mass $3\times10^{7}\,M_\odot$ \cite{2024Natur.628...57F}. Gravitational lensing allows such intrinsically faint accreting BHs to be studied at the epoch of reionization. Similarly, JWST-NIRSpec PRISM spectroscopy characterized GHZ9 at $z=10.145$ to be nitrogen enriched, carbon, and metal poor, with an estimated BH mass of $(1.60 \pm 0.31)\times 10^8\,M_{\odot}$. It is currently the farthest detection by the Chandra X-ray Observatory \cite{2025ApJ...989...75N}. JWST-NIRSpec-IFU observed at $z = 7.15$ has evidence for an accreting BH, traced by a broad component of H${\beta}$ emission \cite{2024MNRAS.531..355U}. The current observational data set therefore combines two qualitatively different populations, the wide-field optical, and near-infrared surveys. Together these surveys sample a substantially broader range of SMBH luminosities and masses, which provide a substantially larger dynamic range, offering a useful test of models in which the formation of early SMBH seeds is tied to the assembly history of their host DM halos.\\

In order to draw parallels between our predicted SMBH mass, with those gathered from observations, in figure \ref{fig:CompMvsz}, we plot the predicted mass of SMBHs within the $16^{\rm th}$ and $84^{\rm th}$ percentile bands, as they evolve with redshift. The yellow and blue regions are representative of the predicted SMBHs, for the benchmark values $\sigma_0$ = 50 and 90 $\rm cm^2/gm$, respectively, with $\omega=200\,\rm km/s$. We plot the observed SMBHs mass with errorbars, alongside the shaded regions, as a function of redshift. We find that our core-collapse scenario accommodates the observed SMBHs inside the $16^{\rm th}$ and $84^{\rm th}$ percentile bands, to a reasonable accuracy. Expectedly, a larger value of $\sigma_0$ is able to explain the more massive SMBHs, observed at early redshifts. This overlap between the theoretical prediction and observed data is seen to increase for higher Eddington accretion rates, explaining the formation of SMBHs of mass as large as $10^{7}\,M_{\odot}$ at redshifts of $\sim10$. The SIDM cross-section is responsible for massive BH seed formation, whereas their growth to SMBHs is determined by mergers and accretion. Our study suggests that a significant fraction of SMBHs either started out from massive seeds or grew at high Eddington accretion rates (i.e. $\eta \geq 1$), at high redshifts. This can be understood from equation \eqref{eq:tcc}, where a larger self-interaction cross-section implies faster core-collapse, hence early formation of SMBH seeds. These early seeds therefore find sufficient time ($t_{\rm obs}-t_{\rm seed}$) to accrete the surrounding gas, under the specified Eddington accretion rates. The left, middle and right panels denote an Eddington accretion rate of 0.6, 1.0, and 1.4, respectively. Observations usually indicate at $\eta \sim 1$ for massive BH seeds \cite{2019ApJ...880...77O,Bogdan:2023ilu,2023ApJ...957L...7K,2024Natur.628...57F}, whereas super-Eddington accretions are realized for low-mass seeds in gas-rich environments \cite{Maiolino:2023zdu,2025ApJ...989...75N}. As discussed earlier, the initial DM cosmology i.e. the epoch at which a core forms, and progenitor halo masses play a role in deciding $M_{\rm seed}$ and $z_{\rm seed}$. To quantify the agreement between the predicted and observed SMBH mass-redshift distributions, we construct a likelihood in logarithmic BH mass. For each observed SMBH $j$, with $y_j=\log_{10}(M_{{\rm SMBH},j}^{\rm obs}/M_\odot)$ and uncertainty $\sigma_j$, the model prediction 	$y_j^{\rm th}(\vec{{\theta}})$ is evaluated at the corresponding observed redshift. At a fixed accretion prescription, the model parameters are $\vec{{\theta}}=(\sigma_0,\omega)$, and we define the reduced $\chi$-squared as,
\begin{equation}
\chi_r^2(\vec{{\theta}}) = \frac{1}{N_{\rm obs}-N_{\theta}}\sum_{j=1}^{N_{\rm obs}} \frac{ \left[ y_j-y_j^{\rm th}(\vec{{\theta}}) \right]^2 }{\sigma_j^2 }.
\label{eq:chisq}
\end{equation}
Here $N_{\rm obs}$ is the number of observed data points, $N_{\theta}=2$.
\begin{figure*}[t!]
	\begin{center}
		\subfloat{\includegraphics[scale=0.354]{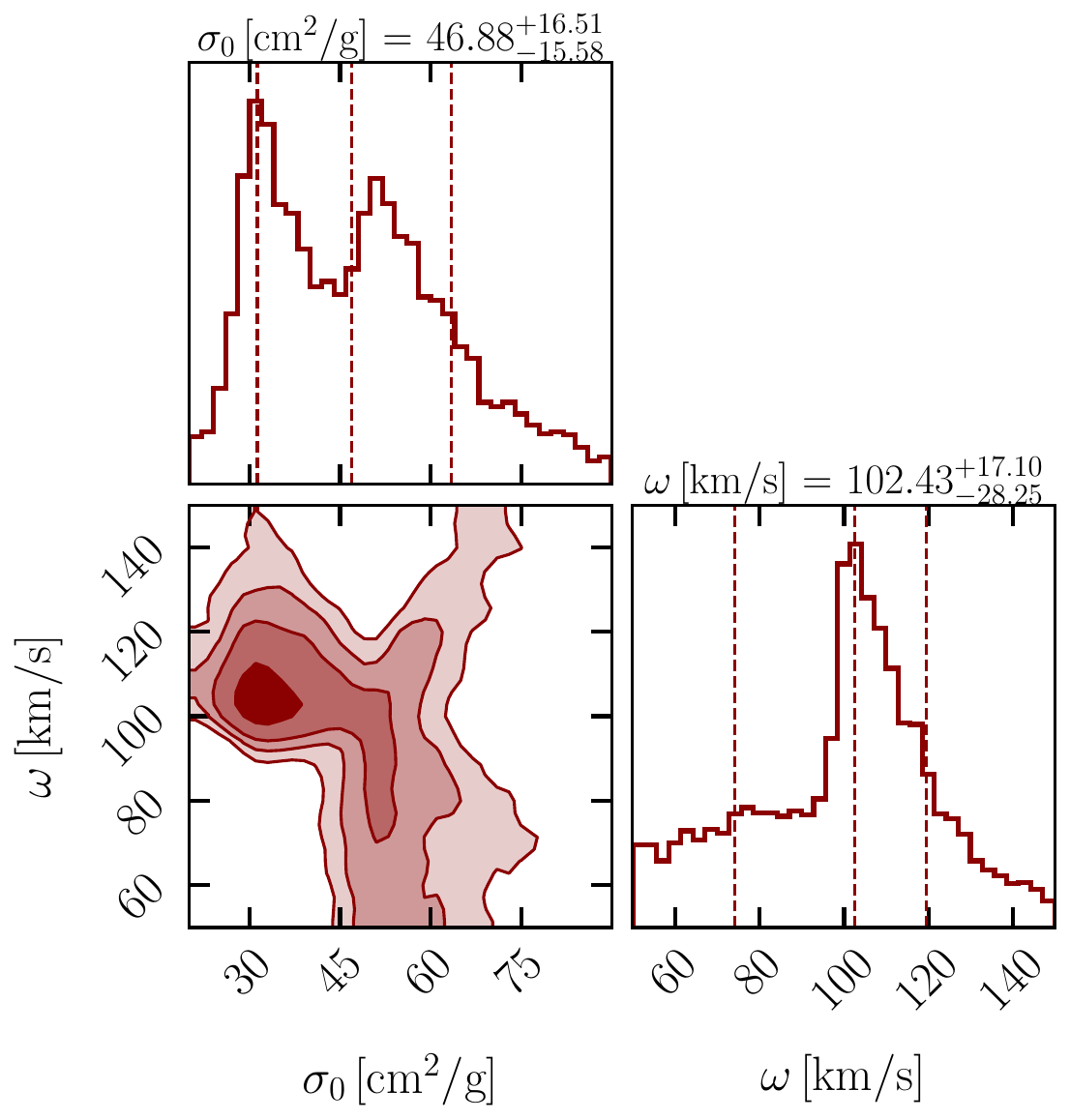}}
		\subfloat{\includegraphics[scale=0.354]{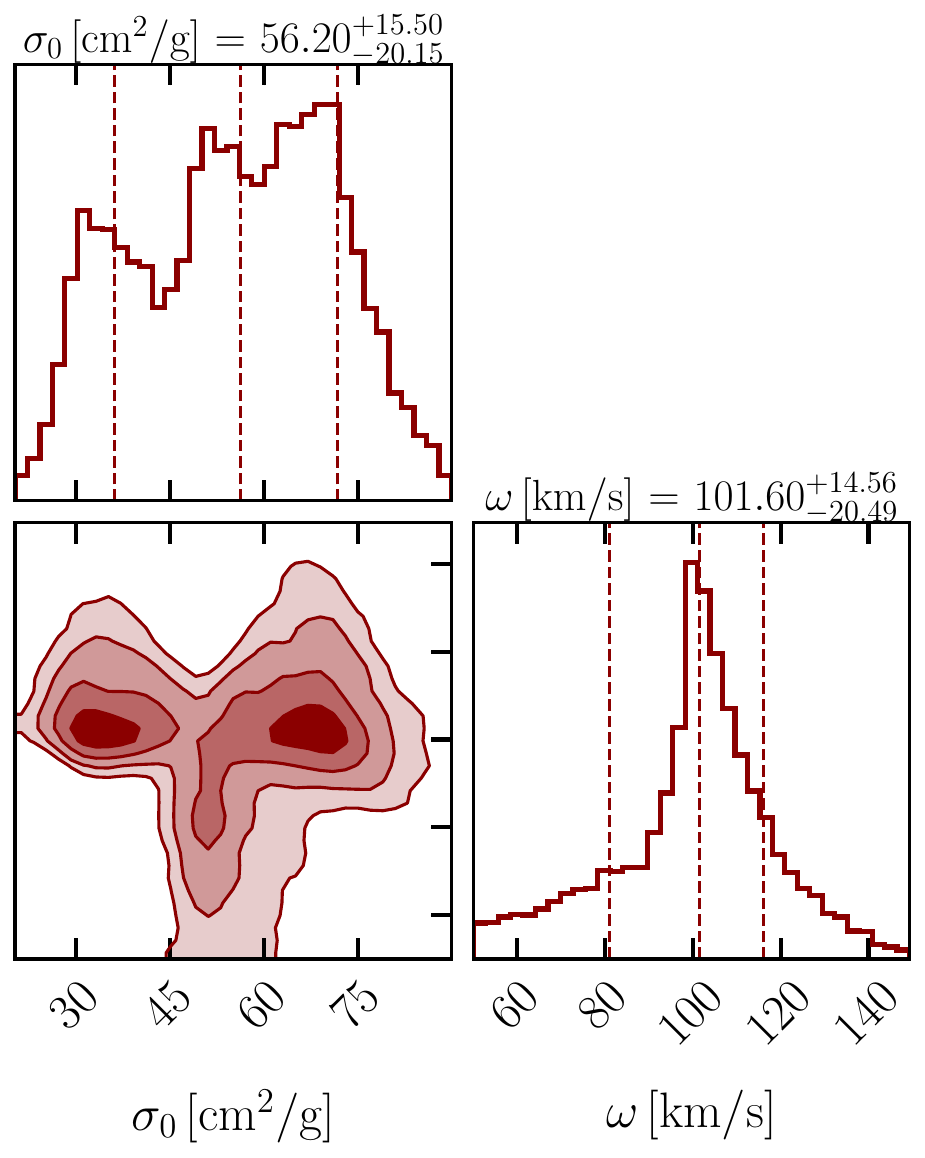}}
		\subfloat{\includegraphics[scale=0.354]{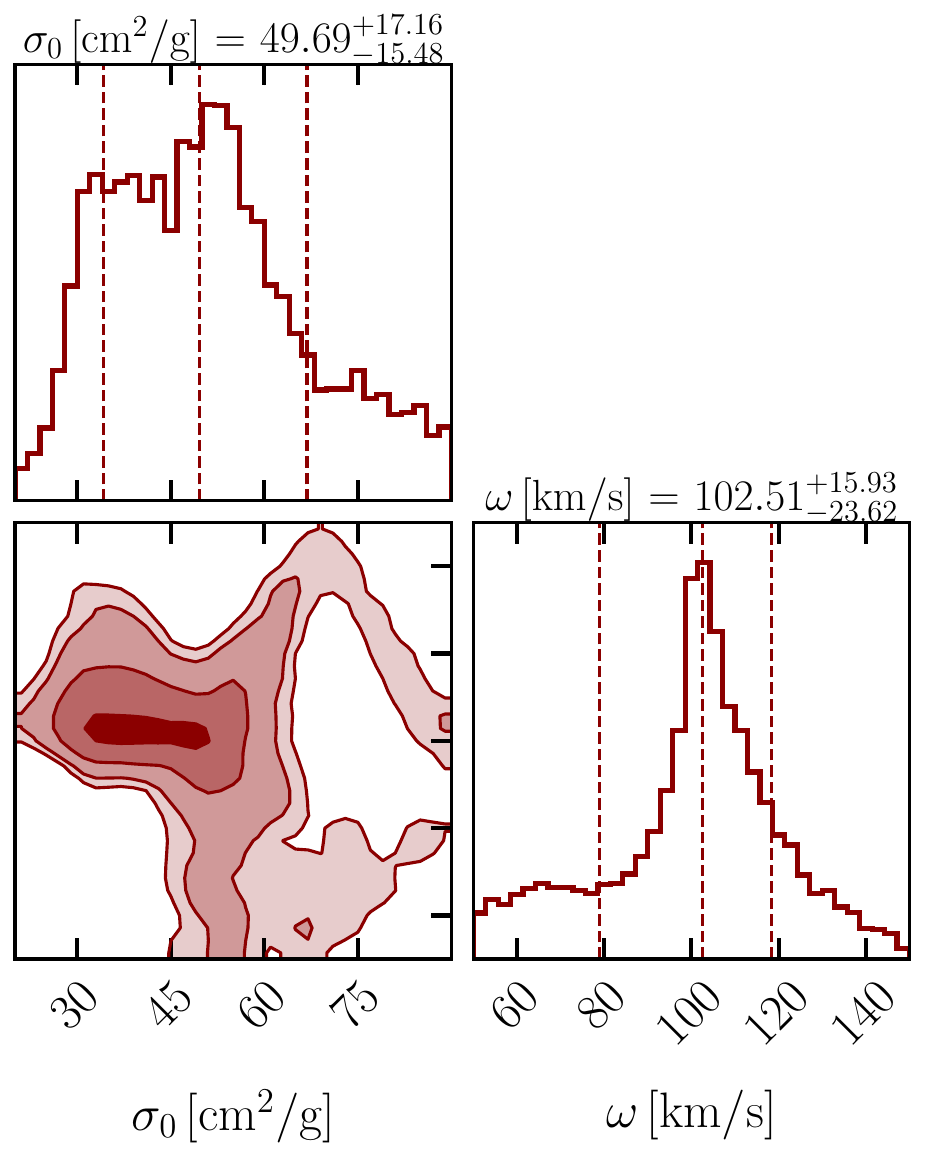}}
		\caption{Contour plots of the posterior distribution functions, for SIDM parameters $\sigma_0/m$ and $\omega$, showing the best-fitted median values with corresponding uncertainties, with Eddington accretion rates 0.6, 1.0, and 1.4 in the left, middle, and right panels, respectively.}
		\label{fig:cornerplot1}
	\end{center}
\end{figure*}
In figure \ref{fig:Chisq}, we plot $\chi^2_r$ following equation \eqref{eq:chisq}, as a function of the DM halo mass in the top panels, and as a function of $\sigma_0$ in the bottom panels, respectively, at redshift 4. Even with different Eddington accretion rates, we find that a higher magnitude of SIDM cross-section results in an increased likelihood of SMBH formation within lower DM halo masses, as can be seen by the green curves in the top panels. Whereas lower $\sigma_0$ results in a preference of higher mass halos for hosting a significant fraction of the SMBH population, as seen from the blue curves in the top panels. These results can be further strengthened by examining the $\chi^2_r$ variation with $\sigma_0$ in the bottom panels of figure \ref{fig:Chisq}, for certain benchmark values of DM halo mass and characteristic velocity scale $\omega$. The blue and green curves display the goodness-of-fit quantified by $\chi_r^2$, for host-halo with mass $10^{10}\,M_{\odot}$ and $10^{11}\,M_{\odot}$ respectively, to host SMBHs with increase in DM self-interaction cross-sections. {This behavior reflects the reduction of gravothermal collapse timescale with increasing interaction strength, thereby extending the efficient seeding regime toward low-mass DM halos. In order to make this discussion more evident, we run a Markov Chain Monte Carlo (MCMC) sampler \cite{Speagle:2019ffr}, using the publicly available python module \texttt{emcee} \cite{2013PASP..125..306F}. In figure \ref{fig:cornerplot1}, we plot the posterior distribution of the SIDM model parameters ($\sigma_0$ and $\omega$). The marginalized posterior distributions shown in figure \ref{fig:cornerplot1} exhibit a relatively stable preference for the characteristic velocity scale $\omega$ across the three accretion prescriptions. For sub-Eddington rates $\eta=0.6$, we obtain $ \sigma_0/m_\chi = 46.88^{+16.51}_{-15.58}\,{\rm cm^2/gm},\,\omega=102.43^{+17.10}_{-28.25}\,{\rm km/s}$. For the Eddington case, $\eta=1.0$, the corresponding posterior constraints become $\sigma_0/m_\chi=56.20^{+15.50}_{-20.15}\,{\rm cm^2/gm},\,\omega=101.60^{+14.56}_{-20.49}\,{\rm km/s}$, while for the super-Eddington limit $\eta=1.4$, we find $\sigma_0/m_\chi=49.69^{+17.16}_{-15.48}\,{\rm cm^2/gm},\, \omega= 102.51^{+15.93}_{-23.62}\,{\rm km/s}$. Here the quoted central values denote the marginalized posterior medians and the uncertainties correspond to the $16^{\rm th}$ to $84^{\rm th}$ percentile Bayesian credible intervals. For the three benchmark accretion rates, and for the population of observed SMBHs, the posterior distributions favors the median values of $\sigma_0$ in the approximate range of $(46-56)\,\rm cm^2/gm$, with $\omega\sim 102\,\rm km/s$.\\
\subsection{Binned SMBH mass function}
\label{subsec:results2}

\begin{figure*}[t!]
	\begin{center}
		\subfloat[\label{sf:margin1}]{\includegraphics[scale=0.27]{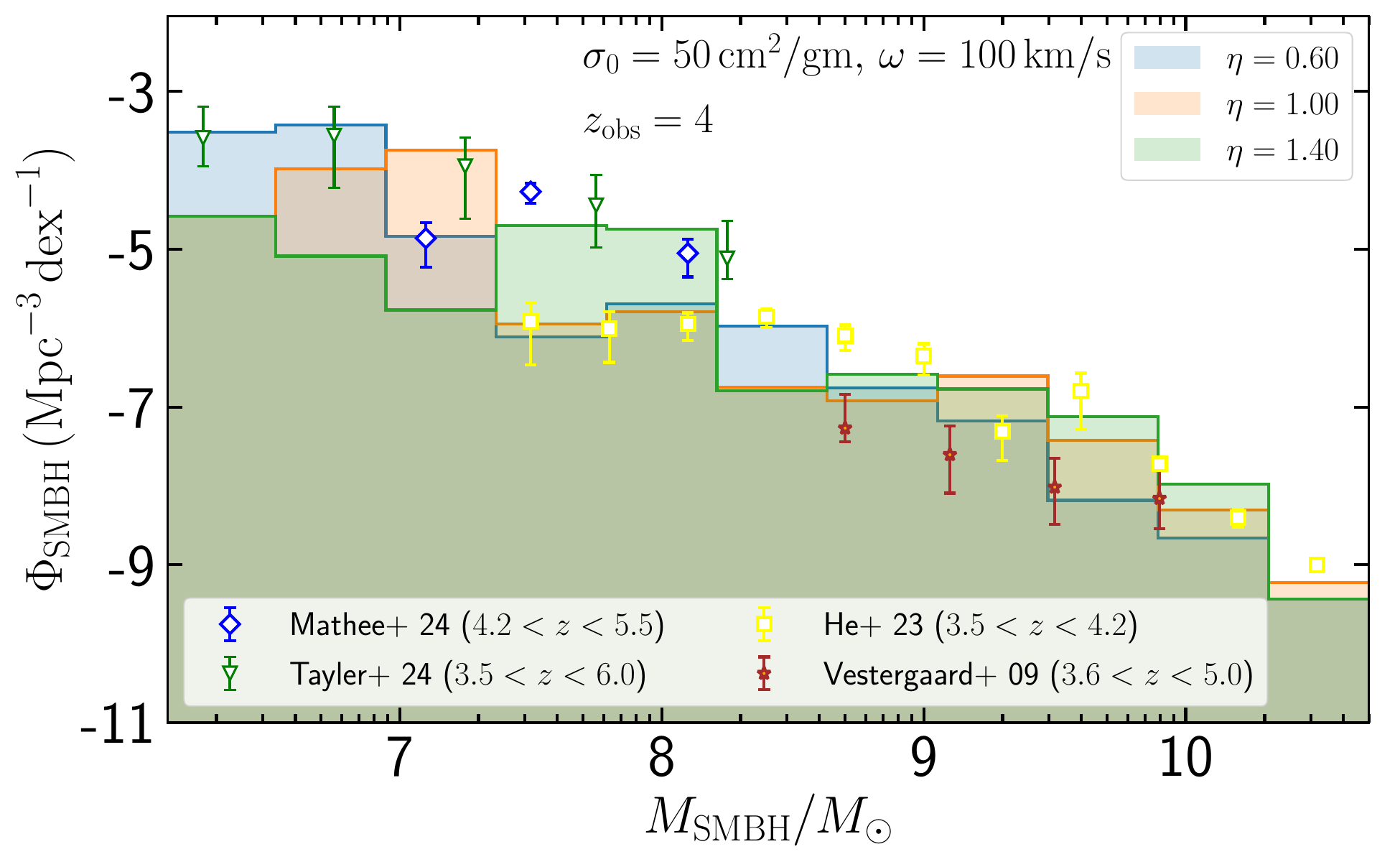}}
		\subfloat[\label{sf:margin2}]{\includegraphics[scale=0.27]{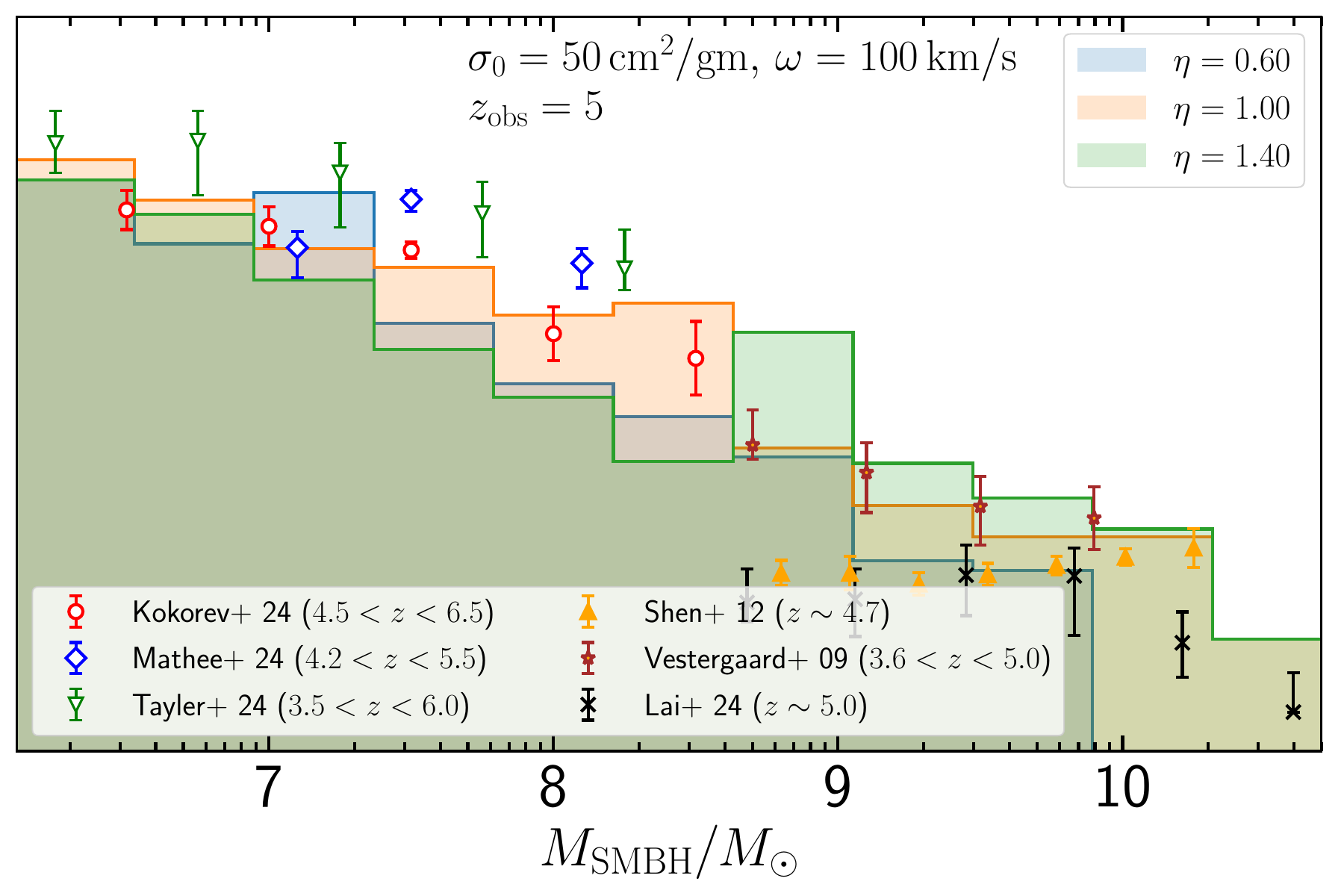}}\\
		\subfloat[\label{sf:margin3}]{\includegraphics[scale=0.27]{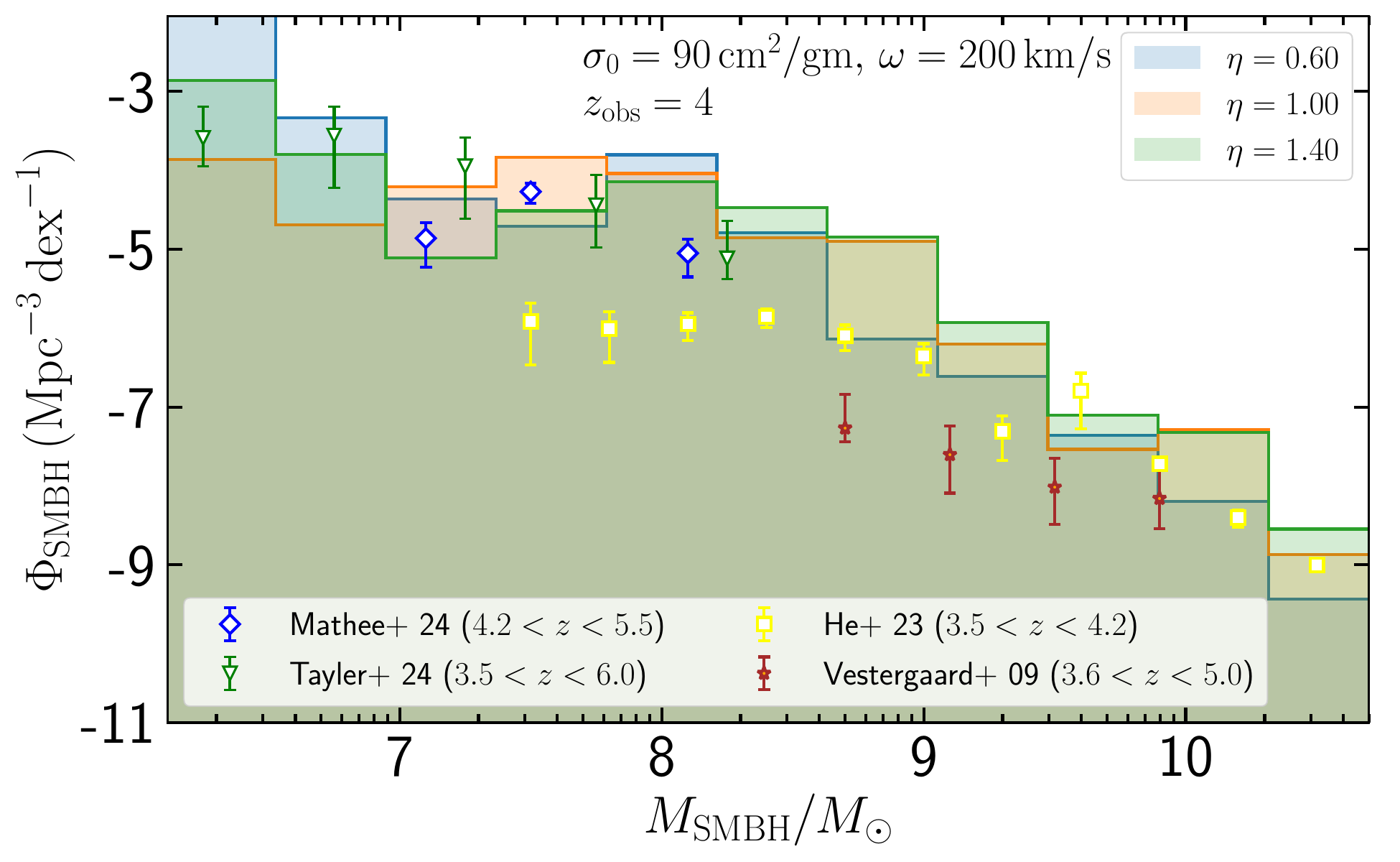}}
		\subfloat[\label{sf:margin4}]{\includegraphics[scale=0.27]{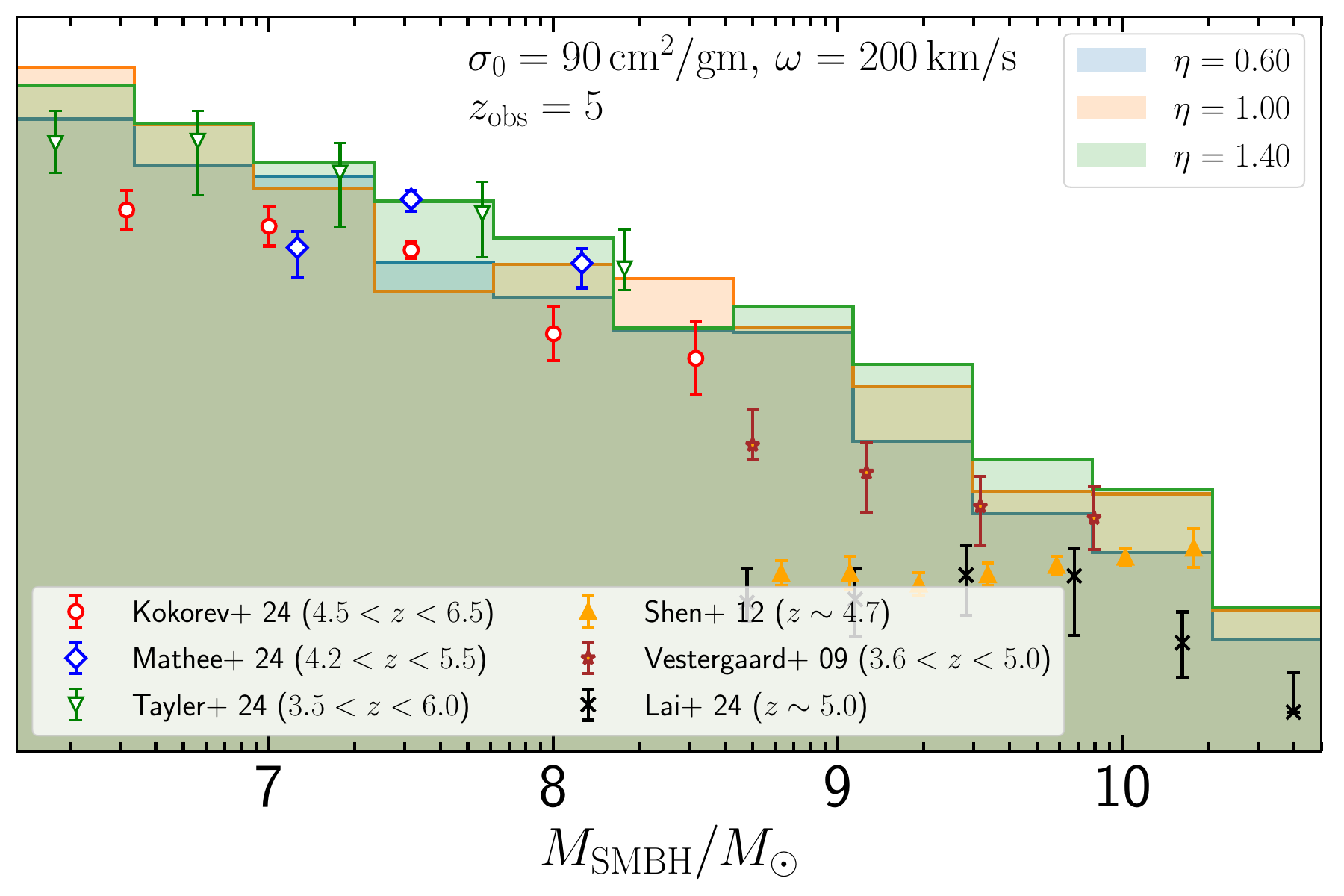}}
		\caption{The binned SMBH mass function for two benchmark SIDM parameters, $\sigma_0/m=50\, \rm cm^2/gm$, $\omega=100\,\rm km/s$, and $\sigma_0/m=90\,\rm cm^2/gm$, $\omega=200\,\rm km/s$, shown in the top and bottom panels, respectively. The left and right panels denote the SMBH mass function at $z$=4, and 5, respectively. The blue, yellow, and green histograms represent Eddington accretion rates of 0.6, 1.0, and 1.4, respectively. The observed SMBH mass functions at specific redshift range are also shown with their corresponding errorbars \cite{2009ApJ...699..800V,2012ApJ...746..169S,Matthee:2023utn,Wu:2022njo,2024MNRAS.531.2245L,2024ApJ...962..152H,2024ApJ...968...38K,2025ApJ...986..165T}.}
		\label{fig:BinnedMF}
	\end{center}
\end{figure*}
The binned SMBH mass function (BHMF) defined as $\Phi_{\rm BH}(M_{\rm BH},z)$, is an estimate of the comoving SMBH number density, evaluated at a given redshift. The BHMF can be estimated by counting the number of BHs in uniform logarithmic mass bins and weighting the individual objects by their survey volume \cite{Kelly:2008fn,Schulze:2014qqa}. Before JWST, high-redshift BHMF measurements were dominated by luminous quasars selected in wide-area optical surveys. The BHMF over a redshift range of $1<z<4.5$ was obtained using the BLR virial masses \cite{2010ApJ...719.1315K,2013ApJ...764...45K}. These studies showed a strong decline in the comoving abundance of massive, and active BHs from $z\gtrsim3$ onward. The high-mass regime $M_{\rm BH}\gtrsim10^{8}\,M_\odot$ has been reliably sampled, while the low-mass BHMF is difficult to constrain because of flux limitations \cite{2010ApJ...719.1315K,2013ApJ...764...45K}. At $z\simeq6$, \cite{2010AJ....140..546W} constructed one of the first high-redshift BHMFs using quasars from the CFHQS survey. They inferred active BHMF covering approximately $M_{\rm BH}\sim10^{8}-10^{10}\,M_\odot$, and showed a declining abundance toward larger $M_{\rm BH}$. JWST measurements have extended direct BHMF towards $M_{\rm BH}\sim10^{6}-10^{8}\,M_\odot$ \cite{2024ApJ...968...38K}. From measurements of $20$ broad-H$\alpha$ AGN at $4.2<z<5.5$, \cite{Matthee:2023utn} inferred the BH masses at $\sim10^{7}-10^{8}\,M_\odot$, and derived the BHMF which indicates a substantially abundant population of moderate-mass active BHs. \cite{2025ApJ...986..165T} inferred BH masses at $M_{\rm BH}\sim10^{6}-10^{8}\,M_\odot$, and constructed a binned BHMF from $50$ H$\alpha$-detected BLR AGN, selected from the CEERS and RUBIES surveys. The analysis includes observational and line-detection completeness corrections, and extends the measured BHMF below $M_{\rm BH}\sim{10}^7\,M_\odot$. The resulting BHMF is approximately power-law like over the measured range, and is broadly consistent with other JWST determinations \cite{2025ApJ...986..165T}. The observed broad-line BHMF is related to the total BHMF through the relation $\Phi_{\rm BH}^{\rm active} \simeq f_{\rm duty}(M_{\rm BH},z)\,\Phi_{\rm BH}^{\rm total}(M_{\rm BH},z)$, where $f_{\rm duty}$ is the active fraction of SMBH or the duty cycle \cite{2010A&A...516A..87S}. Therefore, a theoretical prediction for the total SMBH abundance may not be compared directly with the observed broad-line BHMF, without modeling or assuming the corresponding active BHMF fraction. In figure \ref{fig:BinnedMF}, we plot histograms of the binned BHMF for the SIDM cross-section $\sigma_0/m=50\, \rm cm^2/gm$ and $\omega=100\,\rm km/s$, at redshifts 4 and 5, derived as the best-fit values from our discussions in the previous subsection \ref{subsec:results1}. Additionally, we also plot the binned BHMF, for a set of higher SIDM parameters $\sigma_0/m=90\, \rm cm^2/gm$ and $\omega=200\,\rm km/s$. We derive the theoretical BHMF from our generated sample of SMBHs, shown by the blue, orange, and green bands for an Eddington accretion rate of 0.6, 1.0, and 1.4, respectively. The theoretical BHMF is constructed by combining the SMBH population obtained from our Monte-Carlo merger trees in \texttt{SatGen}, with the cosmological abundance of their host DM halos. For the $i^{\rm th}$ DM halo mass bin containing $N_i$ simulated trees, and spanning across a logarithmic width $\Delta\log_{10}M_{{\rm h}}$, each realization is assigned the cosmological weight
\begin{equation}
w_i =
\frac{1}{N_i}
\left(\frac{dn_{\rm h}}{d\log_{10}M_{\rm h}}\right)_i \Delta\log_{10}M_{{\rm h}},
\end{equation}
where $dn_{\rm h}/d\log_{10}M_{\rm h}$ is the DM halo mass function evaluated at the observation redshift $z_{\rm obs}$ using \texttt{hmf}. The weight $w_i$ therefore represents the comoving number density associated with each simulated halo realization. After evolving the BH seeds through accretion, mergers and SIDM dynamics up to $z_{\rm obs}$, the SMBH mass function within the $\Delta\log_{10}M_{{\rm BH}}$ logarithmic SMBH mass bin is
\begin{equation}
\Phi_{\rm SMBH}(M_{{\rm BH}},z_{\rm obs}) \equiv \frac{dn_{\rm SMBH}}{d\log_{10}M_{\rm BH}}
=
\sum_{i} \frac{N_i^{\rm BH} (z)}{\Delta\log_{10}M_{{\rm BH}}}  w_{i},
\end{equation}
with units of ${\rm Mpc}^{-3}\,{\rm dex}^{-1}$. Here, the sum includes all SMBHs whose final masses fall within the $i^{\rm th}$ host-halo mass bin. $N_i^{\rm BH} (z)$ is the number of SMBHs within the $\Delta \mathrm{log}_{10} M_{\rm BH}$ bin at the target redshift $z_{\rm obs}$ \cite{Shen:2025evo}. Adapting the sampling strategy of \cite{Jiang:2025jtr} to our target redshift at $z=4$, the number of merger tree realizations increase towards lower halo masses in proportion to the halo mass function, taking $N=32$ for the most massive halo bin. This enhanced sampling of low mass halos becomes necessary for generating statistically significant SMBHs through SIDM core collapse. We plot our derived histograms of width 0.4 dex with the observed active SMBH mass functions, available in literature \cite{2009ApJ...699..800V,2012ApJ...746..169S,Matthee:2023utn,Wu:2022njo,2024MNRAS.531.2245L,2024ApJ...962..152H,2024ApJ...968...38K,2025ApJ...986..165T}. For the value of $\sigma_0$, and $\omega$ motivated from our discussion in the previous subsection and from figure \ref{fig:cornerplot1}, we find our predicted SMBH mass functions to overlap with the observed fraction of active SMBH with reasonable agreement, for $\eta \sim (0.6-1.0)$. A higher Eddington accretion rate $\eta \sim 1.4$, over-predicts the observed SMBH mass functions, which gets further amplified for larger SIDM parameters.
\section{Summary and conclusion}
\label{sec:conc}

Self-interactions between DM have been shown to be instrumental in addressing the small-scale discrepancies in structure formation, arising from the $\Lambda$CDM cosmology. DM self-interactions impact the thermalization of galactic cores, influence the abundance and internal structure of satellite galaxies, and alter the perturbation within stellar streams. Self-interaction cross sections typically $\sigma/m_{\chi}\sim\mathcal{O}(1)\,{\rm cm^2/gm}$ are often sufficient to redress these issues to a large extent. Recent observations concerning the diversity in galactic rotation curves of DM-rich environments, such as dwarf and low surface brightness galaxies, and the presence of moderate DM spikes around the central galactic SMBHs, has been argued to prefer larger values of SIDM cross-sections typically $\sigma/m_{\chi}\sim \mathcal{O}(10)\,{\rm cm^2/gm}$, when the interactions are velocity-dependent. In such scenarios, the characteristic velocity scale $\omega$ suppresses the interactions at sufficiently large halo velocities, allowing larger self-interactions in dwarf-scale or satellite halos, while remaining compatible with the constraints from larger astrophysical scales. Thereby accounting for self-interactions across a variety of astrophysical objects. At such high values of DM self-interaction cross-sections, the SIDM halo cores can undergo gravothermal collapse. Such collapsed cores provide a natural mechanism for the early onset of BH formation inside DM halos, and can serve as seeds for the high-redshift SMBH population observed by the JWST and pre-JWST era instruments. \\

In our semi-analytic treatment, SIDM core collapse can produce SMBHs of mass $10^{10}\,M_{\odot}$, by redshifts of 7, hinting at massive seeds with of approximately $M_{\rm seed}\sim10^{7}\,M_{\odot}$. These seeds subsequently grow through accretion with $\eta\simeq0.6-1.4$, and through hierarchical BH mergers, thereby populating the mass and redshift range occupied by the high-redshift SMBHs. The plethora of existing observations therefore provide us the opportunity to probe the cosmological origins of SMBHs from the core-collapse of SIDM cores. We construct an extensive ensemble of halo assembly histories using \texttt{SatGen}, tracing progenitor halos to redshifts as high as $z\simeq20$, and connect these realizations to their cosmological abundances using the halo mass function computed with \texttt{hmf} at $z_{\rm obs} =4$ and 5. We consider host halos spanning the range $M_{\rm halo}=(10^7-10^{12})\,M_{\odot}$, and explore velocity-dependent SIDM models over $\sigma_0=(10-100)\,\rm cm^2/gm$ and $\omega=(50-300)\,\rm km/s$. For each merger tree, we follow the redshift evolution of the halo mass and concentration, evaluate the onset of gravothermal collapse, determine the seed-formation epoch, seed mass, and subsequently evolve the BHs through accretion and hierarchical mergers. The merger treatment further allows us to determine whether BHs formed in satellite progenitors can reach and merge with more massive BHs within the available cosmic time. An important outcome of this analysis is that successful SIDM-induced seed formation constitutes a subset of the halo population. The predicted SMBH abundance is therefore controlled not only by the cosmological halo mass function but also by the probability of sampling halos with sufficiently favorable formation histories and concentrations. A statistical comparison of the predicted and observed SMBH mass-redshift distributions enables us to identify the ranges of DM host-halo masses, and SIDM parameters most conducive to early BH formation. We find that larger values of $\sigma_0$ increase the likelihood of BH formation in lower-mass halos, by reducing the core-collapse timescales. The characteristic core size $r_{1}$, collapse timescale $t_{\rm cc}$, halo concentration $c_{200}$, and halo assembly history therefore act together in determining the occupation probability and mass spectrum of the resulting SMBHs. Comparing our theoretical paradigm to the SMBH observations, we arrive at an approximate range for median values of $\sigma_0 \sim (46-56)\,\rm cm^2/gm$, $\omega \sim 100\,\rm km/s$, and Eddington accretion rates $\eta \sim 0.6-1.4$, to more likely describe the formation of high-redshift SMBHs. We also find the SIDM model parameters $\sigma_0 \sim 50\,\rm cm^2/gm$, $\omega \sim 100\,\rm km/s$ to account for the observed binned SMBH mass functions at redshifts of 4, and 5, with a 100\% duty cycle. The simultaneous agreement with both the SMBH mass-redshift distribution, and the SMBH mass function provides a stronger validation of the SIDM seeding scenario. The former primarily constraining the growth histories, whereas the latter directly probing the cosmological abundance and occupation fraction of BHs. The results we arrive at from our independent analysis can be broadly compared to those existing in literature, derived for SMBH studies.\\

Changes in $\sigma_{0}$ and $\omega$ modify the accessible host-halo mass range, and probability of core collapse, whereas the assumed accretion efficiency controls the subsequent growth and observable abundance of the SMBHs. Therefore, our results on the SIDM parameter space should be interpreted jointly with the uncertainties in BH accretion, merger delays, halo concentration statistics, and observational selection effects. Our results demonstrate that the rapidly expanding census of high-redshift SMBHs provides a complementary cosmological probe of SIDM physics and the cosmology of gravothermal collapse in DM halos. The SIDM cosmology can additionally be verified by gravitational waves emitted during the merger of SMBHs residing inside the progenitor halos. This will take into account the SIDM physics arising from the collapse of SIDM core, and also an accurate inclusion of dynamical friction due to the ubiquitous SIDM media. An improved distribution of BHMF particularly at larger redshifts, derived for the observed fraction of active quasars and those derived from simulations can introduce improved likelihood estimates for the SIDM cosmology. The future detection of SMBH mergers in Pulsar Timing Arrays would provide complementary results to the values presented in our analysis, and narrow down the velocity-dependent SIDM parameter space, with possible smoking-gun signatures for self-interactions between DM. 

\paragraph*{Acknowledgments\,:}
SS would like to thank Debajit Bose for the helpful discussions. UKD acknowledges the support from the Anusandhan National Research Foundation (ANRF), Government of India under Grant Reference No. CRG/2023/003769.

\FloatBarrier
\bibliographystyle{JHEP}
\bibliography{bhcorecollapse.bib,bhcorecollapse2.bib}
\end{document}